\documentclass{LMCS}

\def\doi{9(1:07)2013}
\lmcsheading%
{\doi}
{1--32}
{}
{}
{May\phantom.~31, 2012}
{Feb.~28, 2013}
{}

\usepackage{pgf,tikz}
\usetikzlibrary {arrows}
\usetikzlibrary {decorations}
\usetikzlibrary{snakes}
\usepackage{main}
\usepackage{amsmath,amssymb,wasysym,stmaryrd}
\usepackage{enumerate}
\usepackage{hyperref}
\usepackage{xspace}

\makeatletter
\let\c@definition\c@theorem
\let\c@lemma\c@theorem
\let\c@corollary\c@theorem
\let\c@remark\c@theorem
\let\c@example\c@theorem
\let\c@proposition\c@theorem

\definecolor{magenta}{rgb}{0.5,0,0.5}

\begin{document}
\title{On (Subgame Perfect) Secure Equilibrium in Quantitative
  Reachability Games}

\author[T.~Brihaye]{Thomas Brihaye\rsuper a}
\author[V.~Bruy\`ere]{V\'eronique Bruy\`ere\rsuper a}
\author[J.~De~Pril]{Julie De Pril\rsuper a}
\address{{\lsuper a}University of Mons - UMONS,
   Place du Parc 20, 7000 Mons, Belgium}
\email{\{thomas.brihaye, veronique.bruyere, julie.depril\}@umons.ac.be} 
%\thanks{thanks 1, optional.}	%optional
\author[H.~Gimbert]{Hugo Gimbert\rsuper b}
\address{{\lsuper b}LaBRI \& CNRS, Bordeaux, France}
\email{hugo.gimbert@labri.fr}  

\keywords{turn-based quantitative games, subgame perfect equilibrium,
  secure equilibrium, reachability objectives}

\ACMCCS{[{\bf Theory of computation}]: Theory and algorithms for
  application domains---Algorithmic game theory and mechanism
  design---Algorithmic game theory}
\subjclass{D.2.4}

\begin{abstract}
  We study turn-based quantitative multiplayer non zero-sum games
  played on finite graphs with reachability objectives. In such games,
  each player aims at reaching his own goal set of states as soon as
  possible. A previous work on this model showed that Nash equilibria
  (resp. secure equilibria) are guaranteed to exist in the multiplayer
  (resp. two-player) case. The existence of secure equilibria in the
  multiplayer case remained and is still an open problem. In this
  paper, we focus our study on the concept of subgame perfect
  equilibrium, a refinement of Nash equilibrium well-suited in the
  framework of games played on graphs.  We also introduce the new
  concept of subgame perfect secure equilibrium. We prove the
  existence of subgame perfect equilibria (resp. subgame perfect
  secure equilibria) in multiplayer (resp. two-player) quantitative
  reachability games. Moreover, we provide an algorithm deciding the
  existence of secure equilibria in the multiplayer case.
\end{abstract}

\maketitle

\section*{Introduction}

\subsection*{General framework.} 
The construction of correct and efficient computer systems (hardware
or software) is recognized as an extremely difficult task. To support
the design and verification of such systems, mathematical logic,
automata theory~\cite{HU} and more recently
model-checking~\cite{CGP00} have been intensively studied. The
efficiency of the model-checking approach is widely recognized when
applied to systems that can be accurately modeled as a finite-state
automaton. In contrast, the application of these techniques to more
complex systems like embedded systems or distributed systems has been
less successful. This could be partly explained by the following
reasons: classical automata-based models do not faithfully capture the
complex behavior of modern computational systems that are usually
composed of several interacting components, also interacting with an
environment that is only partially under control. One recent trend to
improve the automata models used in the classical approach of
verification is to generalize these models with the more flexible and
mathematically deeper game-theoretic framework~\cite{Na50,OR94}.

The first steps to extend computational models with concepts from game
theory were done with the so-called two-player zero-sum games played
on graphs~\cite{GTW02}. Those games are adequate to model
controller-environment interactions problems~\cite{T95,Th08}. Moves of
player 1 model actions of the controller whereas moves of player 2
model the uncontrollable actions of the environment, and a winning
strategy for player 1 is an abstract form of a control program that
enforces the control objective. However, only purely antagonist
interactions between a controller and a hostile environment can be
modeled in this framework. In order to study more complex systems with
more than two components and objectives that are not necessarily
antagonist, we need multiplayer non zero-sum games. While in zero-sum
games we look for winning or optimal strategies, in non-zero-sum games
we rather try to find relevant notions of equilibria, like the famous
notion of \emph{Nash equilibrium}~\cite{Na50}. The \emph{secure
  equilibrium}~\cite{CJH06} is a more recent concept that is
especially well-suited for assume-guarantee
synthesis~\cite{CH07,CR10}.
 
There is another interesting extension in such games: moving from
qualitative to quantitative objectives. A player has a qualitative
objective if his aim is to enforce some specification (as, for
instance, reaching a certain set of target states of the graph),
whereas a quantitative objective implies that he wants to minimize or
maximize his gain. For example, a player may wish to reach a set of
target states quickly or with a minimal consumption of energy. Until
now, qualitative objectives have been more studied than quantitative
objectives. However, the latter objectives are as much natural as the
former, and so, aught to be considered. Consequently, we investigate
here \emph{equilibria} for \emph{multiplayer non zero-sum} games
played on graphs with \emph{quantitative} objectives. This article
provides some new results in this research direction, in particular it
is another step in the quest for solution concepts well-suited for the
computer-aided synthesis and verification of multi-agent systems.

\subsection*{Our contribution.} 
We study turn-based multiplayer non zero-sum games played on finite
graphs with quantitative reachability objectives, continuing work
initiated in~\cite{BBD10}. In this framework each player aims at
reaching his own goal as soon as possible. In~\cite{BBD10}, among
other results, it has been proved that a finite-memory Nash
(resp. secure) equilibria always exists in multiplayer
(resp. $2$-player) games.

In this paper we consider alternative solution concepts to the
classical notion of Nash equilibria.  In particular, in the present
framework of games on graphs, it is very natural to consider the
notion of \emph{subgame perfect equilibrium} \cite{selten65}: a choice
of strategies is not only required to be a Nash equilibrium from the
initial vertex, but also after every possible initial history of the
game.  Indeed if the initial state or the initial history of the
system is not known, then a robust controller should be subgame
perfect. We introduce a new and even stronger solution concept with
the notion of subgame perfect \emph{secure} equilibrium, which gathers
both the sequential nature of subgame perfect equilibria and the
verification-oriented aspects of secure equilibria.  These different
notions of equilibria are precisely defined in
Section~\ref{sec:prelim}.

In this paper, we address the following problems:
\begin{pbm}\label{prob:existence}
Given a multiplayer quantitative reachability game $\mathcal{G}$, does
there exist a Nash (resp. secure, subgame perfect, subgame perfect
secure) equilibrium in $\mathcal{G}$?
\end{pbm}

\begin{pbm}\label{prob:finitemem}
Given a Nash (resp. secure, subgame perfect, subgame perfect secure)
equilibrium in a multiplayer quantitative reachability game $\mathcal{G}$,
does there exist such an equilibrium with finite memory?
\end{pbm}

These questions have been positively solved by some of the authors
in~\cite{BBD10} for Nash equilibria in multiplayer games, and
for secure equilibria in two-player games. Notice that these problems
and related ones have been investigated a lot in the qualitative
framework (see~\cite{GU08}).

Here we go a step further and establish the following results about
subgame perfect and secure equilibria:
%\vspace{-.5cm}
\begin{iteMize}{$\bullet$}
  \item
  in every multiplayer quantitative reachability game, there exists a
  subgame perfect equilibrium (Theorem~\ref{theo:ex spe}),
  \item 
  in every two-player quantitative reachability game, there exists a
  subgame perfect secure equilibrium (Theorem~\ref{theo:ex spse}),
  \item 
   in every multiplayer quantitative reachability game, one can decide
   whether there exists a secure equilibrium in {\sf ExpSpace}
   (Theorem~\ref{theo:ex se}),
  \item 
  if there exists a secure equilibrium in a multiplayer quantitative
  reachability game, then there exists one that is finite-memory
  (Theorem~\ref{theo:finitemem se}).
\end{iteMize}

\noindent The results in this paper first appeared in the proceedings of FoSSaCS
2012, \cite{BBDG12}. We here provide their complete proofs.

%\vspace{-.5cm}
\subsection*{Related work.}
Several recent papers have considered \emph{two-player zero-sum} games played
on finite graphs with regular objectives enriched by some
\emph{quantitative} aspects. Let us mention some of them: games with
finitary objectives~\cite{CH06}, games with prioritized
  requirements~\cite{AKW08}, request-response games where the
waiting times between the requests and the responses are
minimized~\cite{HTW08,Z09}, and games whose winning conditions are
expressed via quantitative languages~\cite{BCHJ09}.

Other work concerns \emph{qualitative non zero-sum} games. In~\cite{CJH06}
where the notion of secure equilibrium has been introduced, it is
proved that a unique maximal payoff profile of secure equilibria
always exists for two-player non zero-sum games with regular
objectives. In~\cite{GU08}, general criteria ensuring existence of
Nash equilibria and subgame perfect equilibria (resp. secure equilibria)
are provided for multiplayer (resp. $2$-player) games, as well as
complexity results. In~\cite{BBM10}, the existence of Nash equilibria
is studied for timed games with qualitative reachability
objectives. Complexity issues are discussed in~\cite{BBMU11} about Nash
equilibria in multiplayer concurrent games with B\"uchi objectives. 

Finally, let us mention work that combines both \emph{quantitative}
and \emph{non zero-sum} aspects.  In~\cite{BG09}, the authors study
games played on graphs with terminal vertices where quantitative
payoffs are assigned to the players. These games may have cycles but
all the infinite plays form a single outcome (like in chess where
every infinite play is a draw). That paper gives criteria that ensure
the existence of Nash (and subgame perfect) equilibria in pure and
memoryless strategies.  In~\cite{KLST12}, the studied games are played
on priced graphs similar to the ones considered in this article,
however in a concurrent way. In this concurrent framework, Nash
equilibria are not guaranteed to exist anymore. The authors provide an
algorithm to decide existence of Nash equilibria, thanks to a B\"uchi
automaton accepting all Nash equilibria outcomes. The complexity of
some related decision problems is also studied. In \cite{PS09}, the
authors study Muller games on finite graphs where players have a
preference ordering on the sets of the Muller table.  They show that
Nash equilibria always exist for such games, and that it is decidable
whether there exists a subgame perfect equilibrium.  In both cases
they give a procedure to compute an equilibrium strategy profile (when
it exists). In \cite{FKMSSV10} (respectively \cite{PS11}), it is shown
that every multiplayer sequential game has a subgame-perfect
$\epsilon$-equilibrium for every $\epsilon >0$ if the payoff functions
of the players are bounded and lower-semicontinuous (respectively
upper-semicontinuous).  

\subsection*{Organization of the paper.} 
Section~\ref{sec:prelim} is dedicated to definitions. We present the
kinds of games and equilibria that we study in this paper. In
Section~\ref{sec:spe}, we positively solve
Problem~\ref{prob:existence} for subgame perfect equilibria. In
Section~\ref{sec:spse}, this problem is also positively solved for
subgame perfect secure equilibria, but only in the two-player
case. Finally, in Section~\ref{sec:se}, we study
Problems~\ref{prob:existence} and~\ref{prob:finitemem} in the context
of secure equilibria. We partially solve Problems~\ref{prob:existence}
by providing an algorithm that decides the existence of a secure
equilibrium. And we positively solve Problem~\ref{prob:finitemem} for
secure equilibria.

\section{Preliminaries}\label{sec:prelim}
%======================

%This section is mainly inspired by reference . 

\subsection{Games, Strategy Profiles and Equilibria}\label{sub:def}

In this paper, we distinguish between \emph{qualitative} and
\emph{quantitative} games. In a qualitative game, each player has a
qualitative objective, meaning that he wants to guarantee that some
property holds. In this case, his payoff for a play of the game is
either 1 or 0 (the play does or does not satisfy the property,
respectively). On the other hand, in a quantitative game, each player
has a quantitative objective: he aims at minimizing (or maximizing) a
certain value. His payoff for a play can then be a real number or $\pm
\infty$.

We consider here \emph{quantitative} games played on a graph where all
the players have \emph{reachability objectives}.  It means that, given
a certain set of vertices~$\F_i$, each player~$i$ wants to reach one
of these vertices as soon as possible. We recall the basic notions
about these games and we introduce different kinds of equilibria, like
Nash equilibria.  This section is inspired from~\cite{BBD10}.

\begin{defi}\label{def-game}
  An \emph{infinite turn-based multiplayer quantitative reachability
    game} is a tuple $\game$ where
  \begin{iteMize}{$\bullet$}
  \item[\textbullet] $\Pi$ is a finite set of players,
  \item[\textbullet] $\G$ is a finite directed graph where $V$ is the
    set of vertices, $(V_i)_{i \in \Pi}$ is a partition of $V$ into
    the state sets of each player, $E \subseteq V \times V$ is the set
    of edges, such that for all $v \in V$, there exists $v'\in V$ with
    $(v,v') \in E$ (i.e., each vertex has at least one outgoing edge),
    and
  \item[\textbullet] $\F_i \subseteq V$ is the non-empty goal set of
    player~$i$.
  \end{iteMize}
\end{defi}

\noindent From now on, we often use the term \emph{game} to denote a multiplayer
quantitative reachability game according to Definition~\ref{def-game}.

It is often useful to specify an initial vertex $v_0 \in V$ for a game
$\mathcal{G}$. We call the pair $(\mathcal{G},v_0)$ an
\emph{initialized game}.  Sometimes we omit the word ``initialized''
and just talk about games. The game $(\mathcal{G},v_0)$ is played as
follows. A token is first placed on the vertex~$v_0$.  Player~$i$,
such that $v_0 \in V_i$, has to choose one of the outgoing edges
of~$v_0$ and put the token on the vertex~$v_1$ reached when following
this edge. Then, it is the turn of the player who owns~$v_1$. And so
on.

A \emph{play}~$\rho \in V^{\omega}$ (resp. a \emph{history}~$h \in
V^+$) of $(\mathcal{G},v_0)$ is an \emph{infinite} (resp. a
\emph{finite}) path through the graph~$G$ starting from
vertex~$v_0$. Note that a history is always non-empty because it
starts with~$v_0$.  The set~$H \subseteq V^+$ is made up of all the
histories of~$\mathcal{G}$, and for $\ipi$, the set~$H_i$ is the set
of all histories $h \in H $ whose last vertex belongs to $V_i$.

For any play~$\rho=\rho_0\rho_1\ldots$ of $\mathcal{G}$, we define
$\payoff_i(\rho)$ the \emph{cost} of player~$i$ as:
\begin{equation} \payoff_i(\rho) = \left\{
\begin{array}{ll}
  l & \mbox{ if $l$ is the \emph{least} index such that $\rho_l \in
    \F_i$,}\\ + \infty & \mbox{ otherwise.}
\end{array}\right. \label{eq cost}
\end{equation}
We note $\payoff(\rho) = (\payoff_i(\rho))_{\ipi}$ the \emph{cost
  profile} for the play~$\rho$.  Each player~$i$ aims to
\emph{minimize} the cost he has to pay, i.e. reach his goal set as
soon as possible. The cost profile for a history $h$ is defined
similarly.
%Remark that if $\cost_i(v,v')$ equals to zero for all $\ipi$ and
%$(v,v')\in E$, then we get a \emph{qualitative} game where either a
%player wins (cost $0$), or loses (cost $+ \infty$).

 A \emph{prefix} (resp. \emph{proper prefix})~$\alpha$ of a
 history~$h=h_0\ldots h_k$ is a finite sequence $h_0\ldots h_l$, with
 $l\leq k$ (resp. $l<k$), denoted by $\alpha \leq h$ (resp. $\alpha <
 h$).  We similarly consider a prefix~$\alpha$ of a play~$\rho$,
 denoted by~$\alpha < \rho$. The function~$\last$ returns, given a
 history~$h= h_0\ldots h_k$, the last vertex~$h_k$ of $h$, and the
 \emph{length}~$|h|$ of~$h$ is the number~$k$ of its
 \emph{edges}. Note that the length is not defined as the number of
 vertices. Given a play $\rho=\rho_0\rho_1\ldots$, we denote by
 $\rho_{\le l}$ the prefix of $\rho$ of length $l$, i.e. $\rho_{\le l}
 = \rho_0 \rho_1 \ldots \rho_l$.  Similarly, $\rho_{< l} = \rho_0
 \rho_1 \ldots \rho_{l-1}$.

We say that a play~$\rho=\rho_0\rho_1\ldots$ \emph{visits} a set~$S
\subseteq V$ (resp. a vertex~$v \in V$) if there exists $l \in \N$
such that $\rho_l$ is in $S$ (resp. $\rho_l=v$). The same terminology
also stands for a history~$h$. More precisely, we say that~$\rho$
visits a set~$S$ at (resp. before) \emph{depth}~$\depth \in \IN$ if
$\rho_{\depth}$ is in $S$ (resp. if there exists $l \le \depth$ such
that $\rho_{l}$ is in $S$). For any play~$\rho$ we denote by
$\visit(\rho)$ the set of players~$i \in \Pi$ such that $\rho$ visits
$\F_i$. The set~$\visit(h)$ for a history~$h$ is defined similarly.

A \emph{strategy} of player~$i$ in $\mathcal{G}$ is a
function~$\sigma: H_i \to V$ assigning to each history~$h \in H_i$, a
next vertex~$\sigma(h)$ such that $(\last(h),\sigma(h))$ belongs to
$E$. We say that a play~$\rho=\rho_0\rho_1\ldots$ of $\mathcal{G}$ is
\emph{consistent} with a strategy~$\sigma$ of player~$i$ if
$\rho_{k+1}=\sigma(\rho_0\ldots\rho_k)$ for all $k \in \N$ such that
$\rho_k \in V_i$. The same terminology is used for a history~$h$ of
$\mathcal{G}$.  A \emph{strategy profile} of $\mathcal{G}$ is a
tuple~$(\sigma_i)_{\ipi}$ where $\sigma_i$ is a strategy for
player~$i$.  It determines a unique play in the initialized game
$(\mathcal{G},v_0)$ consistent with each strategy~$\sigma_i$, called
the \emph{outcome} of $(\sigma_i)_{\ipi}$ and denoted by $\langle
(\sigma_i)_{\ipi} \rangle_{v_0}$. We write $\sigma_{-j}$ for
$(\sigma_i)_{\ipi \setminus \{j\}}$, the set of strategies $\sigma_i$
for all the players except for player~$j$.

A strategy~$\sigma$ of player~$i$ is \emph{\memoryless} if $\sigma$
depends only on the current vertex, i.e. $\sigma(h)=\sigma(\last(h))$
for all~$h \in H_i$. More generally, $\sigma$ is a \emph{finite-memory
  strategy} if the equivalence relation~$\approx_{\sigma}$ on $H$
defined by $h \approx_{\sigma} h'$ if $\sigma(h\delta) =
\sigma(h'\delta)$ for all~$\delta \in H_i$ has finite index. In other
words, a finite-memory strategy is a strategy that can be implemented
by a finite automaton with output.  A strategy
profile~$(\sigma_i)_{\ipi}$ is called \emph{\memoryless} or
\emph{finite-memory} if each~$\sigma_i$ is a \memoryless\ or a
finite-memory strategy, respectively.

%For a strategy profile~$(\sigma_i)_{\ipi}$ with outcome $\rho$ and a
%strategy~$\sigma_j'$ of player~$j$, we say that \emph{player~$j$
%  deviates from~$\rho$ before} (resp. \emph{after}) \emph{depth
%  $\depth$} if there exists a prefix~$h$ of~$\rho$, consistent with
%$\sigma_j'$, such that $|h| \leq \depth$ (resp. $|h| \geq \depth$), $h
%\in H_j$ and $\sigma_j'(h) \not = \sigma_j(h)$.

For a strategy profile~$(\sigma_i)_{\ipi}$ with outcome $\rho$ and a
strategy~$\sigma_j'$ of player~$j$, we say that \emph{player~$j$
  deviates from~$\rho$} if there exists a prefix~$h$ of~$\rho$,
consistent with $\sigma_j'$, such that $h\in H_j$ and $\sigma_j'(h)
\not = \sigma_j(h)$.

%\fbox{n\'ecessaire de mettre ``deviates before depth...'' ou
%  compr\'ehensible dans preuve lemme~\ref{lem:remove a cycle}?}

%--------------------
% EQUILIBRE DE NASH |
%--------------------

We now introduce different notions of equilibria in the
\emph{quantitative} framework and give several examples to make clear
the presented concepts. We first begin with the definition of
\emph{Nash equilibrium}.

\begin{defi}\label{def:ne}
  A strategy profile~$(\sigma_i)_{\ipi}$ of a game~$(\mathcal{G},v_0)$
  is a \emph{Nash equilibrium} if for every player~$j \in \Pi$ and
  every strategy~$\sigma_j'$ of player~$j$, we have:
  $$\payoff_j(\rho) \leq \payoff_j(\rho')$$ where $\rho = \langle
  (\sigma_i)_{\ipi} \rangle_{v_0}$ and $\rho' = \langle
  \sigma_j',\sigma_{-j} \rangle_{v_0}$.
\end{defi}
This definition means that for all~$j \in \Pi$, player~$j$ has no
incentive to deviate since he cannot strictly decrease his cost when
using~$\sigma_j'$ instead of~$\sigma_j$.  Keeping notations of
Definition~\ref{def:ne} in mind, a strategy~$\sigma_j'$ such that
$\payoff_j(\rho) > \payoff_j(\rho')$ is called a \emph{profitable
  deviation} for player~$j$ w.r.t.~$(\sigma_i)_{\ipi}$. In this case,
either player~$j$ pays an infinite cost for~$\rho$ and a finite cost
for $\rho'$ (i.e. $\rho'$ visits~$\F_j$, but $\rho$ does not), or
player~$j$ pays a finite cost for~$\rho$ and a strictly lower cost
for~$\rho'$ (i.e. $\rho'$ visits~$\F_j$ for the first time earlier
than~$\rho$ does).

%--------------------
% EQUILIBRE DE SURETE |
%--------------------

We now define the concept of \emph{secure equilibrium}\footnote{Our
  definition naturally extends the notion of \emph{secure equilibrium}
  proposed in~\cite{CJH06} to the quantitative framework.}.  We first
need to associate a binary relation~$\prec_j$ on cost profiles with
each player~$j$. Given two cost profiles~$(x_i)_{\ipi}$ and
$(y_i)_{\ipi}$:
\begin{align*}
  (x_i)_{\ipi} \prec_j (y_i)_{\ipi} \quad \text{iff} \quad & 
  \big(x_j > y_j\big) \vee \\
  & \big(x_j=y_j \wedge (\forall \ipi \ x_i \le y_i )
  \wedge (\exists \ipi \ x_i < y_i)\big)\,.
\end{align*}
We then say that \emph{player~$j$ prefers} $(y_i)_{\ipi}$ \emph{to}
$(x_i)_{\ipi}$. In other words, player~$j$ prefers a cost profile to
another one either if he has a strictly lower cost, or if he keeps the
same cost, the other players have a greater cost, and at least one has
a strictly greater cost.

\begin{defi}\label{def:ES}
  A strategy profile~$(\sigma_i)_{\ipi}$ of a game~$(\mathcal{G},v_0)$
  is a \emph{secure equi\-li\-brium} if for every player~$j \in \Pi$,
  there does not exist any strategy~$\sigma_j'$ of player~$j$ such
  that:
  $$\payoff(\rho) \prec_j \payoff(\rho')$$ where $\rho = \langle
  (\sigma_i)_{\ipi} \rangle_{v_0}$ and $\rho' = \langle
  \sigma_j',\sigma_{-j}\rangle_{v_0}$.
\end{defi}
In other words, player~$j$ has no incentive to deviate w.r.t.
relation~$\prec_j$. A strategy~$\sigma_j'$ such that $\payoff(\rho)
\prec_j \payoff(\rho')$ is called a \emph{$\prec_j$-profitable
  deviation} for player~$j$ w.r.t.  $(\sigma_i)_{\ipi}$. Clearly, any
secure equilibrium is a Nash equilibrium. 

In a secure equilibrium, each player tries first to minimize his own
cost, and then to maximize the costs of the other players.  According
to \cite{CJH06}, a secure profile can be seen as a \emph{contract}
between the players which strengthens \emph{cooperation} in the
following sense: any unilateral selfish deviation by one player cannot
put the other players at a disadvantage if they follow the
contract. For more intuition and motivation about secure equilibria,
see \cite{CJH06,CH07,CR10}.

%--------------------
% EQUILIBRE PARFAITS EN SOUS-JEUX |
%--------------------

We now introduce a third type of equilibrium: the \emph{subgame
  perfect equilibrium}.  In this case, a strategy profile is not only
required to be a Nash equilibrium from the initial vertex, but also
after every possible initial history of the game. Before giving the
definition, we introduce the concept of \emph{subgame} and explain
some notations.

Given a game $\mathcal{G}=(\Pi, V, (V_i)_{i \in \Pi}, E, (\F_i)_{i \in
  \Pi})$, an initial vertex $v_0$, and a history $hv$ of
$(\mathcal{G},v_0)$, with $v \in V$ ($h$ might be empty), the
\emph{subgame $(\mathcal{G}|_h,v)$ of $(\mathcal{G},v_0)$ with history
  $hv$} is the game $\mathcal{G}|_h=(\Pi, V, (V_i)_{i \in \Pi}, E,
(\F_i)_{i \in \Pi})$ initialized at $v$ and such that the cost of a
play~$\pi$ of $(\mathcal{G}|_h,v)$ for player~$i$ is given by
$\cost_i(h\pi)$. Notice that the only difference between
$(\mathcal{G},v_0)$ and $(\mathcal{G}|_h,v)$ occurs in the costs of
the plays. The cost for a play in the subgame $(\mathcal{G}|_h,v)$
depends on the considered history $h$ (the goal set $\F_i$ could have
already been visited by~$h$).  Given a strategy $\sigma_i$ for
player~$i$ in $\mathcal{G}$, we define the strategy $\sigma_i|_h$ in
$\mathcal{G}|_h$ by $\sigma_i|_h(h')=\sigma_i(hh')$ for all histories
$h'$ of $(\mathcal{G}|_h,v)$ such that $\last(h') \in V_i$. Let
$\sigma$ be the strategy profile $(\sigma_i)_{\ipi}$, we write
$\sigma|_h$ for $(\sigma_i|_h)_{\ipi}$, and $h\langle \sigma|_h
\rangle_v$ for the play in $(\mathcal{G},v_0)$ with prefix $h$ that is
consistent with $\sigma|_h$ from $v$.

Then, we say that $(\sigma_i|_h)_{\ipi}$ is a Nash equilibrium in the
subgame $(\mathcal{G}|_h,v)$ if for every player~$j \in \Pi$ and every
strategy~$\sigma_j'$ of player~$j$, we have that $\payoff_j(\rho) \leq
\payoff_j(\rho')$, where $\rho = h\langle (\sigma_i|_h)_{\ipi}
\rangle_v$ and $\rho' = h\langle \sigma_j'
|_h,\sigma_{-j}|_h\rangle_v$.  The definition of a secure equilibrium
in $(\mathcal{G}|_h,v)$ is given similarly.

A subgame perfect equilibrium is a strategy profile that is a Nash
equilibrium after every possible history of the game, i.e. in every
subgame.  In particular, a subgame perfect equilibrium is also a Nash
equilibrium.

\begin{defi}\label{def:SPE}
  A strategy profile~$(\sigma_i)_{\ipi}$ of a game~$(\mathcal{G},v_0)$
  is a \emph{subgame perfect equilibrium} if for all histories $hv$ of
  $(\mathcal{G},v_0)$, with $v \in  V$, $(\sigma_i|_h)_{\ipi}$ is a
  Nash equilibrium in the subgame $(\mathcal{G}|_h,v)$.
\end{defi}

%--------------------
% EQUILIBRE SECURE PARFAITS EN SOUS-JEUX |
%--------------------

We now introduce the last kind of equilibrium that we study. It is a
new notion that combines both concepts of subgame perfect equilibrium
and secure equilibrium in the following way.

\begin{defi}\label{def:SPSE}
  A strategy profile~$(\sigma_i)_{\ipi}$ of a game~$(\mathcal{G},v_0)$
  is a \emph{subgame perfect secure equilibrium} if for all histories
  $hv$ of $(\mathcal{G},v_0)$, with $v \in  V$,
  $(\sigma_i|_h)_{\ipi}$ is a secure equilibrium in the subgame
  $(\mathcal{G}|_h,v)$.
\end{defi}
Notice that a subgame perfect secure equilibrium is a secure
equilibrium, as well as a subgame perfect equilibrium.

In order to understand the differences between the various notions of
equilibria, we provide three simple examples of games limited to two
players and to finite trees.

%After, we will investigate examples of two-player games
%played on graphs involving cycles.

\begin{exa}\label{ex:simple}
  Let $\mathcal{G} = (V,V_1,V_2,E,\F_1,\F_2)$ be the two-player game
  depicted in Fig.~\ref{fig:exsimple1}. The vertices of player~$1$
  (resp. $2$) are represented by circles (resp. squares), that is,
  $V_1 = \{A,D,E,F\}$ and $V_2=\{B,C\}$. The initial vertex~$v_0$ is
  $A$.  The vertices of $\F_1$ are shaded whereas the vertices of
  $\F_2$ are doubly circled; thus $\F_1=\{D,F\}$ and $\F_2=\{F\}$ in
  $\mathcal{G}$.  The number $2$ labeling the edge $(B,D)$ is a
  shortcut to indicate that there are two consecutive edges from $B$
  to $D$ (through one intermediate vertex).  We will keep these
  conventions throughout the article.
  
\begin{figure}[h!]
  \null\hfill
  \begin{minipage}[b]{0.32\linewidth}
    \begin{center}
      \begin{tikzpicture}[xscale=1,yscale=1]
        \everymath{\scriptstyle}
        
        \path (0,2) node[draw,circle,inner sep=2pt] (q0) {$A$};
        \path (-1,1) node[draw,rectangle,inner sep=3.5pt] (q00) {$B$};
        \path (1,1) node[draw,rectangle,inner sep=3.5pt] (q01) {$C$};
        \path (-1,0) node[draw,circle,inner sep=2pt,fill=black!20!white]
        (q000) {$D$};
        \path (.5,0) node[draw,circle,inner sep=2pt] (q010) {$E$};
        \path (1.5,0) node[draw,double,circle,inner sep=2pt,
          fill=black!20!white] (q011) {$F$};

  \draw[arrows=->] (q0) -- (q00) ;
  \draw[arrows=->] (q0) -- (q01) ;

  \draw[arrows=->] (q00) -- (q000) node[pos=.5,left]
  {{$2$}}; 
  \draw[arrows=->] (q01) -- (q010) ; 
  \draw[arrows=->] (q01) -- (q011) ;
\end{tikzpicture}

      \caption{Game $\mathcal{G}$.}
      \label{fig:exsimple1}
    \end{center}
  \end{minipage}
  \hfill
  \begin{minipage}[b]{0.32\linewidth}
    \begin{center}
\begin{tikzpicture}[xscale=1,yscale=1]
  \everymath{\scriptstyle}

  \path (0,2) node[draw,circle,inner sep=2pt] (q0) {$A$};
  \path (-1,1) node[draw,rectangle,inner sep=3.5pt] (q00) {$B$};
  \path (1,1) node[draw,rectangle,inner sep=3.5pt] (q01) {$C$};
  \path (-1,0) node[draw,circle,inner sep=2pt,fill=black!20!white] 
  (q000) {$D$};
  \path (.5,0) node[draw,circle,inner sep=2pt] (q010) {$E$};
  \path (1.5,0) node[draw,double,circle,inner sep=2pt,
    fill=black!20!white] (q011) {$F$};

  \draw[arrows=->] (q0) -- (q00) ;
  \draw[arrows=->] (q0) -- (q01) ;

  \draw[arrows=->] (q00) -- (q000) ; 
  \draw[arrows=->] (q01) -- (q010) ; 
  \draw[arrows=->] (q01) -- (q011) ;
\end{tikzpicture}

      \caption{Game $\mathcal{G}'$.}
      \label{fig:exsimple2}
    \end{center}
  \end{minipage}
  \hfill
  \begin{minipage}[b]{0.32\linewidth}
    \begin{center}
\begin{tikzpicture}[xscale=1,yscale=1]
  \everymath{\scriptstyle}

  \path (0,2) node[draw,circle,inner sep=2pt] (q0) {$A$};
  \path (-1,1) node[draw,rectangle,inner sep=3.5pt] (q00) {$B$};
  \path (1,1) node[draw,rectangle,inner sep=3.5pt] (q01) {$C$};
  \path (-1,0) node[draw,circle,inner sep=2pt,fill=black!20!white] 
  (q000) {$D$};
  \path (.5,0) node[draw,double,circle,inner sep=2pt] (q010) {$E$};
  \path (1.5,0) node[draw,double,circle,inner sep=2pt,
    fill=black!20!white] (q011) {$F$};

  \draw[arrows=->] (q0) -- (q00) ;
  \draw[arrows=->] (q0) -- (q01) ;

  \draw[arrows=->] (q00) -- (q000) ; 
  \draw[arrows=->] (q01) -- (q010) ; 
  \draw[arrows=->] (q01) -- (q011) ;
\end{tikzpicture}

      \caption{Game $\mathcal{G}''$.}
      \label{fig:exsimple3}
    \end{center}
  \end{minipage}
  \hfill\null
\end{figure}

  \noindent In the games $\mathcal{G}$, $\mathcal{G'}$ and $\mathcal{G''}$ of
  Fig.~\ref{fig:exsimple1},~\ref{fig:exsimple2}~and~\ref{fig:exsimple3}
  (played on the same graph), we define two strategies $\sigma_1$,
  $\sigma_1'$ of player~$1$ and two stategies $\sigma_2$, $\sigma_2'$
  of player~$2$ in the following way: $\sigma_1(A)=B$,
  $\sigma'_1(A)=C$, $\sigma_2(C)=E$ and $\sigma'_2(C)=F$. 

  In $(\mathcal{G},A)$, one can easily check that the strategy profile
  $(\sigma_1,\sigma_2)$ is a secure equilibrium (and thus a Nash
  equilibrium) with cost profile is $(3,+\infty)$. Such a secure
  equilibrium exists because player~$2$ threatens player~$1$ to go to
  vertex $E$ in the case where vertex $C$ is reached. This threat is
  not credible in this case since by acting this way, player~$2$ gets
  an infinite cost instead of a cost of $2$ (that he could obtain by
  reaching $F$). For this reason, $(\sigma_1,\sigma_2)$ is not a
  subgame perfect equilibrium (and thus not a subgame perfect secure
  equilibrium).
  %Indeed, $(\sigma_1|_h,\sigma_2|_h)$ is not a Nash
  %equilibrium in the subgame $\mathcal{G}|_h$, where $h=AC$. 
  However,
  one can check that the strategy profile $(\sigma'_1,\sigma'_2)$ is a
  subgame perfect secure equilibrium.

  Let us now consider the game $(\mathcal{G}',A)$ depicted in
  Fig.~\ref{fig:exsimple2} (notice that the number~2 has disappeared
  from the edge $(B,D)$, but $\F_1$ and $\F_2$ remain the same). One
  can verify that the strategy profile $(\sigma_1',\sigma_2')$ is a
  subgame perfect equilibrium which is not a secure equilibrium (and
  thus not a subgame perfect secure equilibrium).  A subgame perfect
  secure equilibrium for $(\mathcal{G}',A)$ is given by the strategy
  profile $(\sigma_1,\sigma_2')$.

  Finally, for the game $(\mathcal{G}'',A)$ depicted in
  Fig.~\ref{fig:exsimple3} (where $\F_1=\{D,F\}$ and $\F_2=\{E,F\}$),
  one can check that the strategy profile $(\sigma_1,\sigma_2')$ is
  both a subgame perfect equilibrium and a secure equilibrium. However
  it is not a subgame perfect secure equilibrium. In particular, this
  shows that being a subgame perfect secure equilibrium is \emph{not}
  equivalent to be a subgame perfect equilibrium and a secure
  equilibrium. On the other hand, $(\sigma_1,\sigma_2)$ is a subgame
  perfect secure equilibrium in $(\mathcal{G}'',A)$.
\end{exa}

The general philosophy of our work is to investigate
\emph{interesting} concepts of equilibria in multiplayer quantitative
reachability games. In these games, each player aims at reaching his
goal set as soon as possible. Having that in mind, a play where a goal
set is visited for the first time after \emph{cycles} were no new goal
set is visited does not seem to be a desirable behavior (see the
definition of \emph{unnecessary cycle} below). It appears thus
reasonable to seek equilibrium concepts with outcomes that do not
present this undesirable feature.

\begin{defi}\label{def:unnecessary cycle}
Given a play $\rho=\alpha\beta\tilde{\rho}$ in a game
$(\mathcal{G},v_0)$, such that $\beta$ is non-empty,
$\last(\alpha)=\last(\alpha\beta)$,
$\visit(\alpha)=\visit(\alpha\beta)$ and
$\visit(\alpha)\not=\visit(\rho)$, the cycle $\beta$ is called
an \emph{unnecessary cycle}.
\end{defi}

\begin{exa}\label{ex:cycle}
Let us exhibit an example of this phenomenon on the two-player game
$(\mathcal{G},A)$ depicted in Fig.~\ref{fig:exsimple4} (we use the
same conventions as in Example~\ref{ex:simple}).  For $n>0$, let us
consider the play $A^nB^\omega$. Along this play, the cycles
$A^{n-1}$, for $n>1$, are unnecessary cycles. Indeed, once $\F_1$ is
visited (in $A$), looping $n$ times on $A$ just delays the apparition
of $\F_2$ (in $B$). However, for each $n >0$, one can build a subgame
perfect equilibrium $(\sigma_1^n,\sigma_2)$ whose outcome is
$A^nB^\omega$ and cost profile is $(0,n)$, as follows:
$$
\sigma_1^n(h)=
\begin{cases}
  A & \text{if } h=A^j, \text{ with } j<n, \\
  B & \text{otherwise.}
\end{cases}
$$ 
This allows us to conclude that the notion of subgame perfect
equilibrium does not prevent the existence of outcomes with
unnecessary cycles. We can notice that $(\sigma_1^n,\sigma_2)$ is not
a secure equilibrium, for all $n > 0$.  However, we will see in the
next example that secure equilibria can also allow this kind of
undesirable behaviors.

\begin{figure}[h]
  \null\hfill
  \begin{minipage}[b]{0.48\linewidth}
    \begin{center}
      \begin{tikzpicture}[xscale=1,yscale=1]
  \everymath{\scriptstyle}
        \path (-2,0) node[draw,circle,fill=black!20!white,inner sep=2pt] (q0) {$A$};
        \path (0,0) node[draw,circle,inner sep=2pt,double] (q1) {$B$};
  \path (-2.6,0) node (q3) {};               
  \draw[arrows=->] (q3) -- (q0) ;           
  \draw[arrows=->] (q0) -- (q1) ;
  \draw[arrows=->] (q0) .. controls +(135:1cm) and +(45:1cm) .. (q0);
  \draw[arrows=->] (q1) .. controls +(135:1cm) and +(45:1cm) .. (q1);
\end{tikzpicture}
      \caption{Subgame perfect equilibrium with outcome $A^nB^{\omega}$.}
      \label{fig:exsimple4}
    \end{center}
  \end{minipage}
  \hfill
  \begin{minipage}[b]{0.48\linewidth}
    \begin{center}
\begin{tikzpicture}[xscale=1,yscale=1]
        \everymath{\scriptstyle}     
        \path (-2,0) node[draw,circle,inner sep=2pt] (q0) {$A$};
        \path (0,0) node[draw,rectangle,inner sep=3.5pt] (q1) {$B$};
        \path (2,0) node[draw,circle,double,fill=black!20!white,inner sep=2pt] (q2) {$C$};
  \path (-2.6,0) node (q3) {};               
  \draw[arrows=->] (q3) -- (q0) ;  
  \draw[arrows=->] (q1) -- (q2) ;

  \draw[arrows=->] (q0) .. controls +(20:1.1cm)  .. (q1);
  \draw[arrows=->] (q1) .. controls +(200:1.1cm)  .. (q0);

  \draw[arrows=->] (q0) .. controls +(135:1cm) and +(45:1cm) .. (q0);
  \draw[arrows=->] (q2) .. controls +(135:1cm) and +(45:1cm) .. (q2);
\end{tikzpicture}
      \caption{Secure equilibrium with outcome $A^nBC^{\omega}$.}
      \label{fig:exsimple5}
    \end{center}
  \end{minipage}
  \hfill\null
\end{figure}

\noindent Let us consider the game of Fig.~\ref{fig:exsimple5} initialized at
$A$. For $n>1$, the cycles $A^{n-1}$ are unnecessary along the play
$A^nBC^\omega$.  However, for each $n >0$, we can build a secure
equilibrium $(\sigma_1^n,\sigma_2^n)$ whose outcome is $A^nBC^\omega$
and cost profile is $(n+1,n+1)$, as follows:
$$
\sigma_1^n(h)=
\begin{cases}
  A & \text{if } h=A^j, \text{ with } j<n, \\
  B & \text{otherwise.}
\end{cases}
\quad\ ; \quad
\sigma_2^n(h)=
\begin{cases}
  C & \text{if } h=A^nB,  \\
  A & \text{otherwise.}
\end{cases}
$$ 
For each $n >0$, the fact that $(\sigma_1^n,\sigma_2^n)$ is a secure equilibrium
is based on the following threat of player~$2$ against
player~$1$: player~$2$ pretends that he will only decide to
visit vertex $C$ if player~$1$ has visited vertex $A$ exactly $n$
times. This behavior is not credible since player~$2$'s interest is to
reach vertex $C$ as soon as possible. In other words, we have that
$(\sigma_1^n,\sigma_2^n)$ is not a subgame perfect equilibrium (and
thus not a subgame perfect secure equilibrium). 
\end{exa}

Those examples motivate the introduction of the notion of subgame
 perfect secure equilibrium. We believe that this notion can help in
 avoiding the undesirable behaviors of unnecessary cycles.
More generally, a deeper understanding of the studied equilibria whose
outcomes have unnecessary cycles could be very useful.  A more subtle
example of a three-player game will be discussed in
Example~\ref{ex:se}.

In the sequel, we study and partially solve
Problem~\ref{prob:existence} and Problem~\ref{prob:finitemem}.  The
next three sections contain useful material for the proofs of our
results.

\subsection{Qualitative Two-player Zero-sum Games}%
\label{subsec:2players}

In this section we recall well-known properties of \emph{qualitative}
two-player zero-sum games~\cite{Th08}. They will be useful for our
proofs, especially in the context of deviations of a player with
respect to a strategy profile: we thus face a two-player zero-sum game
where the player who deviates plays against the \emph{coalition} of
the other players.

We first recall the notion of \emph{weak parity} game.

\begin{defi}\label{def-game-2j-sn}
  A \emph{qualitative two-player zero-sum weak parity game} is a tuple
  $\mathcal{G} = (V, V_1, V_2,\linebreak  E, c)$ where
  \begin{iteMize}{$\bullet$}
  \item[\textbullet] $\Gdjsn$ is a finite directed graph 
    where~$V$ is the set of vertices,~$(V_1,V_2)$ is a partition of~$V$
    into the vertex sets of player~1 and player~2,
    and~$E \subseteq V \times V$ is the set of edges, 
  \item[\textbullet] $c: V \to \IN$ is the coloring function.
  \end{iteMize}
  Player~1 (resp. player~2) \emph{wins} a
  play~$\rho=\rho_0\rho_1\ldots \in V^{\omega}$ of the
  game~$\mathcal{G}$ if the maximum color in the sequence $c(\rho_0)
  c(\rho_1) c(\rho_2) \ldots$ is even (resp.~odd).
\end{defi}
Given an initial vertex~$v_0 \in V$, the notions
of \emph{play}, \emph{history} and \emph{strategy} are the same as the
ones defined in Section~\ref{sub:def}.  The game is
said \emph{zero-sum} because every play is won by exactly one of the
two players.

In zero-sum games, it is interesting to know if one of the players can
play in such a way that he is sure to win, however the other player
plays. This is formalized with the notion of
\emph{winning strategy}.  A strategy~$\sigma_i$ for player~$i$ is a
winning strategy \emph{from an initial vertex}~$v$ if all the plays
of~$\mathcal{G}$ starting from~$v$ that are consistent with~$\sigma_i$
are won by player~$i$.  If player~$i$ has a winning strategy
in~$\mathcal{G}$ from~$v$, we say that player~$i$ \emph{wins} the
game~$\mathcal{G}$ from~$v$.  We say that a game~$\mathcal{G}$ is
\emph{determined} if for all~$v \in V$, one of the two players has a
winning strategy from~$v$.

Martin showed \cite{M75} that every qualitative two-player zero-sum
game with a Borel type winning condition is determined. In particular,
we have the following proposition:

\begin{prop}\cite[Theorem~5]{Th08}\label{prop:zerosum} 
Let~$\gamedjsnwparity$ be a qualitative two-player zero-sum weak
parity game.  Then for all~$v \in V$, one of the two players has a
\emph{\memoryless}\ winning strategy from~$v$ (in particular,
$\mathcal{G}$ is determined).
\end{prop}

 We here consider three special cases of the weak parity
condition: \emph{reachability}, \emph{safety}, and
\emph{reachability under safety} conditions. A \emph{qualitative
  two-player zero-sum reachability under safety game} is denoted
$$\mathcal{G}=(V,V_1,V_2,E,R,S)$$ where $R, S \subseteq V$ and
$R \ne \emptyset$.  In such a game, player~$1$ wins a play~$\rho$ iff
$\rho$ visits $R$ (i.e., $\exists i \ \rho_i \in R$) while staying in
$S$ (i.e., $\forall i \ \rho_i \in S$). The reachability under safety
condition can be encoded with a weak parity condition by defining the
coloring function $c$ as follows: $c(v)=3$ if $v
\not\in S$, $c(v)=2$ if $v \in R$ and $c(v)=1$
otherwise. \emph{Reachability} games (resp. \emph{safety} games) are
special cases of reachability under safety games
$\mathcal{G}=(V,V_1,V_2,E,R,S)$ where $S=V$ (resp. $R=V$). We can now
state a corollary of Proposition~\ref{prop:zerosum}.

\begin{cor}\label{coro:gamei}
 Let~$\mathcal{G}=(V,V_1,V_2,E,R,S)$ be a qualitative two-player
 zero-sum reachability under safety game. Then the game $\mathcal{G}$
 is determined and player~$1$ has a \memoryless\ stra\-tegy~$\strat_1$
 that enables him to reach~$R$ within $|V|-1$ edges, while staying in
 $S$, from each vertex~$v$ from which he wins the game.
\end{cor}

In the sequel, we apply Corollary~\ref{coro:gamei} on
particular \emph{two-player} games. Given a multiplayer quantitative
reachability game $\game$ and a player $i \in \Pi$, we denote by
$\mathcal{G}_i = (V,V_i,V\setminus V_i,E,R,S)$ (or $(G_i,R,S)$ in
short) the qualitative two-player zero-sum reachability under safety
game associated with player~$i$. This game is played on the graph
$G_i=(V,V_i,V\setminus V_i,E)$, where player~$i$ plays against the
coalition of all the other players. Player~$i$ controls the vertices
of~$V_i$ and the coalition those of~$V \setminus V_i$; player~$i$ aims
at reaching $R$ while staying in $S$, and the coalition wants to
prevent this.

\subsection{Unraveling}

In the proofs of this article, it will be often useful to
\emph{unravel} the graph $G=(V,(V_i)_{i\in \Pi},E)$ from an
initial vertex~$v_0$, which ends up in an \emph{infinite tree},
denoted by~$\tinf$.  This tree can be seen as a new graph where the
set of vertices is the set~$H$ of histories of~$\mathcal{G}$, the
initial vertex is~$v_0$, and a pair~$(h,hv) \in H \times H$ is an edge
of~$\tinf$ if~$(\last(h),v) \in E$. A history~$h$ is a vertex of
player~$i$ in~$\tinf$ if~$h \in H_i$, and $h$ belongs to the goal set
of player~$i$ if~$\last(h) \in \F_i$.

We denote by~$\ginf$ the related game. This game~$\ginf$ played on the
un\-ra\-ve\-ling~$\tinf$ of~$G$ from~$v_0$ is equivalent to the
game~$(\mathcal{G},v_0)$ played on the graph~$G$ in the following
sense.  A play~$(\rho_0)(\rho_0\rho_1)(\rho_0\rho_1\rho_2)\ldots$
in~$\ginf$ induces a unique play~$\rho = \rho_0\rho_1\rho_2\ldots$
in~$(\mathcal{G},v_0)$, and conversely.  Thus, we denote a play
in~$\ginf$ by the respective play in~$(\mathcal{G},v_0)$. The
bijection between plays of~$(\mathcal{G},v_0)$ and plays of~$\ginf$
allows us to use the same cost function~$\payoff$, and to transform
easily strategies in~$\mathcal{G}$ to strategies in~$\ginf$ (and
conversely).

For practical reasons, we often consider equivalently $\mathcal{T}$ in
our proofs instead of $(\mathcal{G},v_0)$, and the equilibria defined
in $\mathcal{T}$ are obviously equilibria in
$(\mathcal{G},v_0)$. Moreover, figures given in proofs to help the
understanding roughly represent the unraveling $\tinf$ of $G$ and
plays in game $\mathcal{T}$.

 We also need to study the tree~$\tinf$ limited to a certain
 depth~$\depth \in \IN$: we denote by $\tfin$ the \emph{truncated tree
   of $\tinf$ of depth $\depth$} and~$\gfin$ the \emph{finite} game
 played on~$\tfin$.  More precisely, the set of vertices of~$\tfin$ is
 the set of histories~$h \in H$ of length~$\leq\depth$; the edges
 of~$\tfin$ are defined in the same way as for~$\tinf$, except that
 for the histories~$h$ of length~$\depth$, there exists no
 edge~$(h,hv)$.  A play~$\rho$ in~$\gfin$ corresponds to a history
 of~$(\mathcal{G},v_0)$ of length \emph{equal to}~$\depth$.  The
 notions of cost and strategy are defined exactly like in the
 game~$\ginf$, but limited to the depth~$\depth$. For instance, a
 player pays an infinite cost for a play~$\rho$ (of length~$\depth$)
 if his goal set is not visited by~$\rho$.

\subsection{Kuhn's Theorem}

%This section is devoted to the classical Kuhn's theorem
%\cite{kuhn53ega}. It claims the existence of a subgame perfect
%equilibrium (resp. subgame perfect secure equilibrium) in multiplayer
%(resp. two-player) games played on \emph{finite trees}.
This section is devoted to the classical Kuhn's theorem
\cite{kuhn53ega}. It claims the existence of a subgame perfect
equilibrium (resp. subgame perfect secure equilibrium) in multiplayer
games played on \emph{finite trees}.

A \emph{preference relation} is a total, reflexive and transitive
binary relation.

\begin{thm}[Kuhn's theorem]\label{theo:kuhn}
  Let~$\tree$ be a \emph{finite} tree and~$\mathcal{G}$ a game played
  on~$\tree$. For each player~$i \in \Pi$, let~$\precsim_i$ be a
  preference relation on cost profiles.  Then there exists a strategy
  profile~$(\sigma_i)_{\ipi}$ such that for every history $hv$ of
  $\mathcal{G}$, every player~$j \in \Pi$, and every strategy~$\sigma_j'$
  of player~$j$ in~$\mathcal{G}$, we have
  $$\payoff(\rho') \precsim_j \payoff(\rho)$$
  where~$\rho=h \langle (\sigma_i|_h)_{\ipi} \rangle_v$ 
  and~$\rho' =h \langle \sigma_j'|_h,\sigma_{-j}|_h \rangle_v$.
\end{thm}

One can easily be convinced that the binary relation on cost profiles
used to define the notion of Nash equilibrium (see
Definition~\ref{def:ne}) is total, reflexive and transitive. We thus
have the following corollary.

\begin{cor}\label{coro:kuhn Nash}
  Let~$(\mathcal{G},v_0)$ be a game and~$\tinf$ be the unraveling
  of~$G$ from~$v_0$. Let $\gfin$ be the game played on the truncated
  tree of~$\tinf$ of depth~$\depth \in \IN$. Then there exists a
  subgame perfect equilibrium in~$\gfin$.
\end{cor}

%Let $\preceq_j$ be the relation defined by $x \preceq_j y$ iff $x
%\prec_j y$ or $x=y$, where $\prec_j$ is the relation used in
%Definition~\ref{def:ES}. We notice that in the \emph{two-player} case,
%the relation $\preceq_j$ is total, reflexive and transitive. So, in
%the two-player case, Kuhn's theorem stands for subgame perfect secure
%equilibrium.
Let $\preceq_j$ be the relation defined by $x \preceq_j y$ iff $x
\prec_j y$ or $x=y$, where $\prec_j$ is the relation used in
Definition~\ref{def:ES}. We notice that in the two-player case, this
relation is total, reflexive and transitive. However when there are
more than two players, $\preceq_j$ is no longer total. Nevertheless,
it is proved in \cite{Roux09} that Kuhn's theorem remains true when
$\preceq_j$ is only transitive. So, the next corollary holds.

\begin{cor}\label{coro:kuhn secure}
  Let~$(\mathcal{G},v_0)$ be a game and~$\tinf$ be the unraveling
  of~$G$ from~$v_0$. Let $\gfin$ be the game played on the truncated
  tree of~$\tinf$ of depth~$\depth \in \IN$. Then there exists a
  subgame perfect secure equilibrium in~$\gfin$.
\end{cor}

%Let us remark that in the multiplayer case when $|\Pi| > 2$, the
%relation $\preceq_j$ is reflexive and transitive, but not
%total. Nevertheless, it has been proved in \cite{Roux09} that Kuhn's
%theorem remains true when the binary relation $\precsim_i$ is only
%transitive.

\section{Existence of a Subgame Perfect Equilibrium}\label{sec:spe}

In this section, we positively solve Problem~\ref{prob:existence} for
subgame perfect equilibria.

% and for subgame perfect secure equilibria but for two-player games
% only.

\begin{thm}\label{theo:ex spe}
  In every multiplayer quantitative reachability game, there exists a
  subgame perfect equilibrium.
\end{thm}

The proof uses techniques completely different from the ones given
in~\cite{BBD10,BBDlong} for the existence of Nash equilibria, and
secure equilibria in two-player games.

Let $(\mathcal{G},v_0)$ be a game and $\mathcal{T}$ be the infinite
game played on the unraveling $T$ of $G$ from $v_0$.
%This existence result is based on the approximation of the
%infinite duration game by games of finite duration.
Kuhn's theorem (and in particular Corollary~\ref{coro:kuhn Nash})
guarantees the existence of a subgame perfect equilibrium in each
finite game $\ttree{n}$ for every depth~$n \in \IN$.  Given a sequence
of such equilibria, the keypoint is to derive the existence of a
subgame perfect equilibrium in the infinite game $\mathcal{T}$. This
is possible by the following lemma.

\begin{lem}\label{lemma:subseq}
Let $(\sigman)_{n \in \IN}$ be a sequence of strategy profiles
such that for every $n\in\IN$,
$\sigman$ is a strategy profile in the truncated game $\ttree{n}$.
Then there exists a strategy profile $\sigmal$ in the game
$\mathcal{T}$
with the property:
\begin{equation}\label{eq:subseq}
\forall \depth \in \IN,~\exists n \geq \depth,
% \text{ such that }
~\sigmal \text{ and } \sigma^{n} \text{ coincide on histories of length
up to }\depth.
\end{equation}
\end{lem}
\proof
%%\begin{proof}
This result is a direct consequence of the compactness of the set of
infinite trees with bounded outdegree~\cite{kechris95}.  An
alternative proof is as follows.  We give a tree structure, denoted by
$\Gamma$, to the set of all strategy profiles in the games
$\ttree{n}$, $n \in \IN$: the nodes of $\Gamma$ are the strategy
profiles, and we draw an edge from a strategy profile $\sigma$ in
$\ttree{n}$ to a strategy profile $\sigma'$ in $\ttree{n+1}$ if and
only if $\sigma$ is the restriction of $\sigma'$ to histories of
length less than $n$. It means that the nodes at depth $d$ correspond
to strategy profiles of $\ttree{d}$. We then consider the tree
$\Gamma'$ derived from $\Gamma$ where we only keep the nodes
$\sigma^{n}$, $n \in \IN$, and their ancestors.  Since $\Gamma'$ has
finite outdegree, it has an infinite path by K{\"o}nig's lemma. This
path goes through infinitely many nodes that are ancestors of nodes in
the set $\{\sigma^n, n \in \IN\}$.  Therefore there exists a strategy
profile $\sigmal$ in the infinite game~$\mathcal{T}$ (given by the previous
infinite path in $\Gamma'$) with property~\eqref{eq:subseq}.
\qed
%%\end{proof}

\proof[Proof of Theorem~\ref{theo:ex spe}]
%%\begin{proof}[of Theorem~\ref{theo:ex spe}] 
Let $\game$ be a multiplayer quantitative reachability game, $v_0$ be
 an initial vertex, and $\ginf$ be the game played on the unraveling
 of $G$ from~$v_0$.  For all $n \in \IN$, we consider the finite game
 $\ttree{n}$ and get a subgame perfect equilibrium $\sigman=
 (\sigman_i)_{\ipi}$ in this game by Corollary~\ref{coro:kuhn
   Nash}. According to Lemma~\ref{lemma:subseq}, there exists a
 strategy profile $\sigmal$ in the game $\mathcal{T}$ with
 property~\eqref{eq:subseq}.
%   Let
%   $\sigmal$ a startegy profile in $\cal T$ with property Then $\sigman$ is extended
%   arbitrarily to the infinite game $\ginf$. So we obtain a sequence
%   $(\sigman)_{n \in \IN}$ of strategy profiles in the game $\ginf$. By
%   compactness of the set of strategy profiles of $\mathcal{G}$ (see
%   \fbox{ref}), there exists a convergent subsequence. We denote it by
%   $(\sigmakn)_{n \in \IN}$ and the limit strategy profile by
%   $\sigmal$.

  It remains to show that $\sigmal$ is a subgame perfect equilibrium
  in $\ginf$, and thus in $(\mathcal{G},v_0)$. Let $hv$ be a history
  of the game (with $v \in V$). We have to prove that $\sigmal|_h$ is
  a Nash equilibrium in the subgame $(\ginf|_h,v)$. As a
  contradiction, suppose that there exists a profitable deviation
  $\sigma'_j$ for some player~$j \in \Pi$ w.r.t. $\sigmal|_h$ in
  $(\ginf|_h,v)$. This means that $\cost_j(\rho) > \cost_j(\rho')$ for
  $\rho=h\langle \sigmal|_h\rangle_v$ and $\rho'=h\langle
  \sigma'_j|_h,\sigmal_{-j}|_h\rangle_v$, that is, $\rho'$ visits $\F_j$
  for the first time at a certain depth $\depth$, such that $|h| < d <
  +\infty$, and $\rho$ visits $\F_j$ at a depth strictly greater than
  $\depth$ (see Figure~\ref{fig:spe}).  Thus:
  % we have
  %that 
  $$\cost_j(\rho) > \cost_j(\rho')=\depth.$$ 

\begin{figure}[h!] 
  \centering
    \begin{tikzpicture}[yscale=.5,xscale=.75]
      \everymath{\scriptstyle}
      \draw (0,8) -- (-5,0);
      \draw (0,8) -- (5,0);

      \draw (-4.5,0.7) node[below] (q0) {$\ginf$};
      \draw (-3.1,0.7) node[below] (q0) {$(\ginf|_h,v)$};
      
      \draw[fill=black] (0,8) circle (1pt);

      \draw[dotted] (-4,2) -- (5.1,2);
     \draw[very thin,dotted] (-4,4) -- (5.1,4);

       \path (5.1,4) node[right] (q0) {$\depth$};%{$\trunc{\depth}(T)$}; 
       \path (5.1,2) node[right] (q0) {$n$};%{$\trunc{k_n}(T)$}; 

      \draw[very thick] (0,8) .. controls (-.2,7.1) .. (.1,6.4);
      \draw[fill=black] (.1,6.4) circle (1.1pt);
      \draw (.05,6.4) node[right] (q0) {$v$};
      \path (0,7.1) node[left] (q0) {$h$};

      \draw (.1,6.4) -- (-4,0);
      \draw (.1,6.4) -- (4.1,0);

      \draw[thick] (.1,6.4) .. controls (.7,3.6) .. (.8,0);

      \path (.7,0) node[below] (q0) {$\rho$};

     \draw[fill=black] (.3,5.5) circle (1pt);

      \draw[thick] (.3,5.5) .. controls (-1,4) .. (-1.7,0);

      \path (-1.7,.1) node[below] (q0) {$\rho'$};

      \draw[fill=black] (.3,5.5) circle (1pt);

      \draw[fill=black] (-.8,4) circle (2pt);

      \path (-.8,4) node[right] (q0) {$\F_j$};

      \draw[fill=black] (-.92,3.7) circle (1pt);
      \draw[thick] (-.92,3.7) .. controls (-1.7,3) .. (-2.1,2);
      \path (-2.1,2.1) node[below] (q0) {$\pi'$};

      \draw[fill=black] (.6,3.8) circle (1pt);
      \draw[thick] (.6,3.8) .. controls (1.6,3.2) .. (1.8,2);
      \path (1.8,2) node[below] (q0) {$\pi$};
      
    \end{tikzpicture}
    \caption{The game $\ginf$ with its subgame $(\ginf|_h,v)$.}
    \label{fig:spe}
  \end{figure}

\noindent According to property~\eqref{eq:subseq}, there exists $n \ge \depth$
such that $\sigmal$ coincide with $\sigman$ on histories of length up
to $d$.
  %for all $k_n
  %\ge k_{n_0}$, we have $$\sigmakn = \sigmal \text{ in } \ttree{\depth}.$$
  %
  %According to property~\eqref{eq:subseq} there exists an integer $k_{n_0}$
  %such that $\sigmal$ and $\sigmakn$
  %for
  %$d:=\max\{d,k_{n_0}\}$, 
  It follows that for $\pi=h\langle \sigman|_h\rangle_v$
  and $\pi'=h\langle \sigma'_j|_h, \sigman_{-j}|_h \rangle_v$,
  we have that 
  (see Figure~\ref{fig:spe})
  $$\cost_j(\pi')=\cost_j(\rho')=\depth \quad \text{and} \quad \cost_j(\pi)
  > d,$$ as $\pi'$ and $\rho'$ coincide up to depth $d$.
  And so, $\sigma'_j$ is a profitable deviation for player~$j$
  w.r.t. $\sigman|_h$ in $(\ttree{n}|_h,v)$, which leads to a
  contradiction with the fact that $\sigman$ is a subgame perfect
  equilibrium in $\ttree{n}$ by hypothesis. 
  \qed
%%\end{proof}

As an extension, we consider \emph{multiplayer quantitative
  reachability games with tuples of costs on edges} (as in
\cite{BBDlong}). In these games, we assume that edges
are labelled with tuples of strictly positive costs (one cost for each
player). Here we do not only count the number of edges to reach the
goal of a player, but we sum up his costs along the path until his
goal is reached. His aim is still to minimize his global cost for a
play. Let us give the formal definition.

\begin{defi}\label{def-game-costs}
  A \emph{multiplayer quantitative reachability game with tuples of
    costs on edges} is a tuple $\gamecosts$ where
  \begin{iteMize}{$\bullet$}
  \item[\textbullet] $\Pi$ is a finite set of players,
  \item[\textbullet] $\G$ is a finite directed graph where $V$ is the
    set of vertices, $(V_i)_{i \in \Pi}$ is a partition of $V$ into
    the state sets of each player, and $E \subseteq V \times V$ is the
    set of edges, such that for all $v \in V$, there exists $v'\in V$
    with $(v,v') \in E$,
  \item[\textbullet] $\cost_i: E \to \R^{>0}$ is the cost function of
    player~$i$ defined on the edges of the graph,
  \item[\textbullet] $\F_i \subseteq V$ is the non-empty goal set of
    player~$i$.
  \end{iteMize}
\end{defi}

\noindent In this context, we adapt the definition of $\payoff_i(\rho)$, the
\emph{cost} of player~$i$ for a play~$\rho=\rho_0\rho_1\ldots$\,:
\begin{equation} \payoff_i(\rho) = \left\{
\begin{array}{ll}
  \displaystyle\sum_{k=1}^{l} \cost_i((\rho_{k-1},\rho_k)) &
  \mbox{ if $l$ is the \emph{least} index such that $\rho_l \in
    \F_i$,}\\
%\mbox{ where $l=\ind_i(\rho)$,}\\
  + \infty & \mbox{ otherwise.} 
\end{array}\right. \label{eq cost tuple}
\end{equation}
%Remark that in Section~\ref{sec:prelim}, we defined $\payoff_i(\rho)$ by
%$\ind_i(\rho)$ if $\rho$ visits $\F_i$.
In this framework, we also prove the existence of a subgame perfect
equilibrium. The proof is similar to the one of Theorem~\ref{theo:ex
  spe}, the only difference lies in the choice of the considered
depth~$\depth$.

\begin{thm}\label{theo:ex spe tuple of costs}
  In every multiplayer quantitative reachability game with tuples of
  costs on edges, there exists a subgame perfect equilibrium.
\end{thm}

Let us introduce some notations that will be useful for the proof of
this theorem. We define $\cmin := \min_{\ipi} \min_{e \in E}
\cost_i(e)$, $\cmax := \max_{\ipi} \max_{e \in E} \cost_i(e)$ and
$\K:= \left\lceil \frac{\cmax}{\cmin} \right\rceil$. It is clear that
$\cmin, \cmax >0$ and $\K \geq 1$.

\proof[Proof of Theorem~\ref{theo:ex spe tuple of costs}]
%%\begin{proof}[of Theorem~\ref{theo:ex spe tuple of costs}] 
Let $\gamecosts$ be a multiplayer quantitative reachability game with
tuples of costs on edges, $v_0$ be an initial vertex, and $\ginf$ be
the game played on the unraveling of $G$ from~$v_0$.  For all $n \in
\IN$, we consider the finite game $\ttree{n}$ and get a subgame
perfect equilibrium $\sigman= (\sigman_i)_{\ipi}$ in this game by
Corollary~\ref{coro:kuhn Nash}. According to Lemma~\ref{lemma:subseq},
there exists a strategy profile $\sigmal$ in the game $\mathcal{T}$
with property~\eqref{eq:subseq}.

  We then show that $\sigmal$ is a subgame perfect equilibrium in
  $\ginf$, and thus in $(\mathcal{G},v_0)$. Let $hv$ be a history of
  the game ($v \in V$). We have to prove that $\sigmal|_h$ is a Nash
  equilibrium in the subgame $(\ginf|_h,v)$. As a contradiction,
  suppose that there exists a profitable deviation $\sigma'_j$ for
  some player~$j \in \Pi$ w.r.t. $\sigmal|_h$ in $(\ginf|_h,v)$. This
  means that $\cost_j(\rho) > \cost_j(\rho')$ for $\rho=h\langle
  \sigmal|_h\rangle_v$ and $\rho'=h\langle
  \sigma'_j|_h,\sigmal_{-j}|_h\rangle_v$. Thus $\rho'$ visits $\F_j$ for
  the first time at a certain depth $\depth'$, such that $|h| <
  \depth' < +\infty$.

  We define some depth~$\depth$ depending on the fact that $\rho$ visits
  $\F_j$ or not.
  \[ \depth = \left\{
  \begin{array}{ll}
    \max\{\depth',\depth''\} & \mbox{ if $\rho$ visits $\F_j$ for the
      first time at depth~$\depth''$,}\\ 
    \depth' \cdot \K & \mbox{ if $\rho$ does not visit $\F_j$.}
  \end{array}\right. \]  
  According to property~\eqref{eq:subseq}, there exists $n \ge \depth$
  such that $\sigmal$ coincide with $\sigman$ on histories of length
  up to $\depth$.  For $\pi=h\langle
  \sigman|_h\rangle_v$ and $\pi'=h\langle \sigma'_j|_h, \sigman_{-j}|_h
  \rangle_v$, since $\depth \ge \depth'$, it follows that:
  $$\cost_j(\pi')=\cost_j(\rho').$$
 
  If $\rho$ visits $\F_j$, then it holds that
  $\cost_j(\pi)=\cost_j(\rho)$ by definition of $\depth$, and so
  $\cost_j(\pi) > \cost_j(\pi')$. If $\rho$ does not visit $\F_j$,
  then the following inequalities hold:
  $$\cost_j(\pi') \le \depth' \cdot \cmax \le \depth \cdot \cmin <
  \cost_j(\pi).$$ The first inequality comes from the fact that $\pi'$
  visits $\F_j$ at depth $\depth'$, the second one from the definition
  of $\depth$, and the last one from the fact that if $\pi$ visits
  $\F_j$, it must happen after depth $\depth$ (as $\rho$ does not visit
  $\F_j$).

  In both cases $\cost_j(\pi)>\cost_j(\pi')$, and we conclude that
  $\sigma'_j$ is a profitable deviation for player~$j$
  w.r.t. $\sigman|_h$ in $(\ttree{n}|_h,v)$, which leads to a
  contradiction with the fact that $\sigman$ is a subgame perfect
  equilibrium in $\ttree{n}$ by hypothesis.  
  \qed
%%\end{proof}

\begin{rem}
  We can transform the cost functions $(\cost_i)_{\ipi}$ (\eqref{eq
    cost} or \eqref{eq cost tuple}) of our games in the following way:
  for any player~$i$ and any play~$\rho$,
  \[\cost_i'(\rho) = \left\{
    \begin{array}{ll}
      1-\frac{1}{c+1} & \mbox{\ if $\cost_i(\rho)=c \in {\mathbb R}^+$,}\\ 
      1 & \mbox{\ if $\cost_i(\rho)=+\infty$.}
    \end{array}\right. \]
  These new cost functions $(\cost_i')_{\ipi}$ are bounded and
  continuous (in the product topology on $V^\omega$).  Moreover, a
  subgame perfect equilibrium in a game with the cost functions
  $(\cost_i)_{\ipi}$ is a subgame perfect equilibrium in this game
  with the new cost functions $(\cost_i')_{\ipi}$, and
  conversely. Then, Theorems~\ref{theo:ex spe} and~\ref{theo:ex spe
    tuple of costs} are consequences of~\cite{har85,FL83}.

\end{rem}

\section{Existence of a Subgame Perfect Secure Equilibrium}\label{sec:spse}
% in the Two-player Case}

Regarding subgame perfect \emph{secure} equilibria, we positively
solve Problem~\ref{prob:existence} but only in the case of two-player
games. 

\begin{thm}\label{theo:ex spse}
  In every two-player quantitative reachability game, there exists a
  subgame perfect secure equilibrium.
\end{thm}
The main ideas of the proof are similar to the ones for
Theorem~\ref{theo:ex spe}.

\proof[Proof of Theorem~\ref{theo:ex spse}]
%%\begin{proof}[of Theorem~\ref{theo:ex spse}] 
Let $\gamedeux$ be a two-player quantitative reachability game, $v_0$
be an initial vertex, and $\ginf$ be the game played on the unraveling
of $G$ from~$v_0$.  For every $n \in \IN$, we consider the finite game
$\ttree{n}$ and get a subgame perfect secure equilibrium $\sigman=
(\sigman_1,\sigman_2)$ in this game by Corollary~\ref{coro:kuhn
  secure}.  According to Lemma~\ref{lemma:subseq} there exists a
strategy profile $\sigmal$ in the game $\mathcal{T}$ such that
$\sigmal$ has property~\eqref{eq:subseq}.
  
  %By compactness of the set of strategy profiles of
  %$\mathcal{G}$, there exists a convergent
  %subsequence. We denote it by $(\sigmakn)_{n \in \IN}$ and the limit
  %strategy profile by $\sigmal$.

  We show that $\sigmal=(\sigmal_1,\sigmal_2)$ is a subgame perfect
  secure equilibrium in $\ginf$. Let $hv$ be a history of the game ($v
  \in V$). We have to prove that $\sigmal|_h$ is a secure equilibrium
  in the subgame $(\ginf|_h,v)$. As a contradiction,
  suppose that there exists a $\prec_j$-profitable deviation
  $\sigma'_j$ for some player~$j \in \{1,2\}$ w.r.t. $\sigmal|_h$ in
  $(\ginf|_h,v)$. Let us assume w.l.o.g. that $j=1$. As $\sigmal|_h$ is a
  Nash equilibrium in $(\ginf|_h,v)$ (see the proof of
  Theorem~\ref{theo:ex spe}), we know that
  \begin{equation}\label{eq:dev}  
  \cost_1(\rho)=\cost_1(\rho') \text{ and } \cost_2(\rho) <
  \cost_2(\rho')
  \end{equation} 
  where $\rho = h\langle \sigmal_1|_h, \sigmal_2|_h\rangle_v$ and $\rho'
  = h\langle \sigma'_1|_h, \sigmal_2|_h\rangle_v$. Thus it implies that
  $\cost_2(\rho)$ is finite. Let $\depth$ be the maximum between
  $\cost_1(\rho)$ and $\cost_2(\rho)$ if $\cost_1(\rho)$ is finite, or
  $\cost_2(\rho)$ otherwise. Remark that $d > |h|$.
  According to property~\eqref{eq:subseq}, there exists $n \ge d$ 
  such that the strategy profiles $\sigmal$ and $\sigman$ coincide on
  histories of length up to $d$.
  % By
  %Lemma~\ref{lemma:subseq}, there exists $k_{n_0} \in \IN$ such that
  %for all $k_n \ge k_{n_0}$, we have
  %\begin{equation}\label{eq:cvg}  
  %  \sigmakn = \sigmal \text{ in } \ttree{\depth}.
  %\end{equation} 
  %With $k_n:=\max\{d,k_{n_0}\}$, 

  Let us show that $\sigma'_1$ would then be a $\prec_1$-profitable
  deviation for player~$1$ w.r.t. $\sigman|_h$ in $(\ttree{n}|_h,v)$.  
  In this aim we first prove that 
  \begin{equation}\label{eq:cost2} \cost_2(\pi)< \cost_2(\pi') \end{equation}
  where $\pi = h\langle \sigman_1|_h, \sigman_2|_h\rangle_v$ and $\pi' =
  h\langle \sigma'_1|_h, \sigman_2|_h\rangle_v$ are finite plays in
  $\ttree{n}$ (see Fig.~\ref{fig:spse}). 
  By definition of $\depth$ and according to
  property~\eqref{eq:subseq}, we have that
  $\cost_2(\pi)=\cost_2(\rho) \le \depth$. If
  $\cost_2(\rho')=\cost_2(\pi')$, Equation~\eqref{eq:dev} implies that
  $\cost_2(\pi) < \cost_2(\pi')$. Otherwise, we have that
  $\cost_2(\pi') > \depth$ as $\rho'$ and $\pi'$ coincide until depth
  $\depth$ (by property~\eqref{eq:subseq}), and then
  $\cost_2(\pi) \le \depth < \cost_2(\pi')$.

\begin{figure}[h!] 
  \centering
    \begin{tikzpicture}[yscale=.5,xscale=.75]
      \everymath{\scriptstyle}
      \draw (0,8) -- (-5,0);
      \draw (0,8) -- (5,0);

      \draw (-4.5,0.7) node[below] (q0) {$\ginf$};
      \draw (-3.1,0.7) node[below] (q0) {$(\ginf|_h,v)$};
      
      \draw[fill=black] (0,8) circle (1pt);

      \draw[dotted] (-4,2) -- (5.1,2);
     \draw[very thin,dotted] (-4,4) -- (5.1,4);

       \path (5.1,4) node[right] (q0) {$\depth$};%{$\trunc{\depth}(T)$}; 
       \path (5.1,2) node[right] (q0) {$n$};%{$\trunc{k_n}(T)$}; 

      \draw[very thick] (0,8) .. controls (-.2,7.1) .. (.1,6.4);
      \draw[fill=black] (.1,6.4) circle (1.1pt);
      \draw (.05,6.4) node[right] (q0) {$v$};
      \path (0,7.1) node[left] (q0) {$h$};

      \draw (.1,6.4) -- (-4,0);
      \draw (.1,6.4) -- (4.1,0);

      \draw[thick] (.1,6.4) .. controls (.7,3.6) .. (.8,0);

      \path (.7,0) node[below] (q0) {$\rho$};

      \draw[fill=black] (.53,4.3) circle (2pt);     
      \path (.47,4.3) node[right] (q0) {\small$\F_2$};

     \draw[fill=black] (.3,5.5) circle (1pt);

      \draw[thick] (.3,5.5) .. controls (-1,4) .. (-1.7,0);

      \path (-1.7,.1) node[below] (q0) {$\rho'$};

      \draw[fill=black] (.3,5.5) circle (1pt);

      \draw[fill=black] (-.92,3.7) circle (1pt);
      \draw[thick] (-.92,3.7) .. controls (-1.7,3) .. (-2.1,2);
      \path (-2.1,2.1) node[below] (q0) {$\pi'$};

      \draw[fill=black] (.6,3.8) circle (1pt);
      \draw[thick] (.6,3.8) .. controls (1.6,3.2) .. (1.8,2);
      \path (1.8,2) node[below] (q0) {$\pi$};
      
    \end{tikzpicture}
    \caption{The game $\ginf$ with its subgame $(\ginf|_h,v)$.}
    \label{fig:spse}
  \end{figure}
  
  \noindent We now consider $\cost_1(\pi)$ and $\cost_1(\pi')$. Let us study the
  next two cases.  
  \begin{iteMize}{$\bullet$}
  \item If $\cost_1(\rho) < +\infty$,
  then we have that 
  \begin{equation}\label{eq:cost1egal} 
  \cost_1(\pi) = \cost_1(\pi') 
  \end{equation} 
  because $\cost_1(\rho') = \cost_1(\rho) = \cost_1(\pi)
  = \cost_1(\pi') \le \depth$ by Equation~\eqref{eq:dev},
  property~\eqref{eq:subseq} and definition of $\depth$. 
  \item If $\cost_1(\rho) = +\infty$, then we show that
  $\cost_1(\pi)=+\infty$, and as a consequence we get
  that 
  \begin{equation}\label{eq:cost1ge} 
  \cost_1(\pi) \ge \cost_1(\pi').  
  \end{equation}
  As a contradiction suppose that $\cost_1(\pi)<+\infty$. Consider
  vertex $\rho_{\depth}$, the first vertex of $\rho$ that belongs to
  $\F_2$ (we recall that $\cost_2(\rho) = d$).  Suppose that
  player~$1$ has a winning strategy to reach his goal from vertex
  $\rho_d$ in the zero-sum reachability game $\mathcal{G}_1 =
  (G_1,\F_1,V)$ (as defined in Section~\ref{subsec:2players}). Then
  this contradicts the fact that $\sigmal$ is a subgame perfect
  equilibrium in $\ginf$ (see the proof of Theorem~\ref{theo:ex
    spe}). Therefore, by determinacy of $\mathcal{G}_1$
  (Corollary~\ref{coro:gamei}), player~$2$ has a winning strategy from
  vertex $\rho_d$ to prevent player~$1$ from reaching $\F_1$. But in
  this case, this strategy is a $\prec_2$-profitable deviation
  w.r.t. $\sigman|_h$ in $(\ttree{n}|_h,v)$, because player~$2$ can
  keep his cost while strictly increasing player~$1$'s cost. This is
  impossible as $\sigman$ is a subgame perfect secure equilibrium in
  $\ttree{n}$. Thus, we must have that $\cost_1(\pi)=+\infty$.
  \end{iteMize} 
  In all possible situations, we proved that $\sigma'_1$ is a
  $\prec_1$-profitable deviation for player~$1$ w.r.t. $\sigman|_h$ in
  $(\ttree{n}|_h,v)$ because either $\cost_1(\pi)=\cost_1(\pi')$ and
  $\cost_2(\pi) < \cost_2(\pi')$, or $\cost_1(\pi) > \cost_1(\pi')$
  (see (\ref{eq:cost2}--\ref{eq:cost1ge})).  So we get a contradiction
  with the fact that $\sigman$ is a subgame perfect secure equilibrium
  in $\ttree{n}$ by hypothesis.  
\qed
%%\end{proof}

%Unfortunately the proof does not seem
%to extend to the multiplayer case. Indeed we face the same kind
%of problems encountered in~\cite{BBDlong} when trying to establish the
%existence of secure equilibria in the multiplayer case
%(see~\cite{BBDlong} for further discussion on these problems). 
Unfortunately the proof does not seem to extend to the multiplayer
case. Indeed we face the same kind of problems encountered
in~\cite{BBD10,BBDlong}, where the existence of secure equilibria is
proved for two-player games and left open for multiplayer
games.
%we consider \emph{simple} cost functions
%$(\cost_i)_{\ipi}$ such that for all $i$, $j \in \Pi$ we have that
%$\cost_i=\cost_j$ and $\cost_i:E \to \N_0$).

\section{Decidability of the Existence of a Secure Equilibrium}%
\label{sec:se}

In this section, we study Problems~\ref{prob:existence}
and~\ref{prob:finitemem} in the context of secure equilibria. Both
problems have been positively solved in~\cite{BBD10} for
two-player games only.  To the best of our knowledge, the existence of
secure equilibria in the multiplayer framework is still an open
problem. We here provide an algorithm that \emph{decides} the
existence of a secure equilibrium.  We also
show that if there exists a secure equilibrium, then there exists one
that is finite-memory.
 
\begin{thm}\label{theo:ex se}
  In every multiplayer quantitative reachability game, one can decide
  whether there exists a secure equilibrium in {\sf ExpSpace}.
\end{thm}

\begin{thm}\label{theo:finitemem se}
If there exists a secure equilibrium in a multiplayer quantitative 
reachability game, then there exists one that is finite-memory.
\end{thm}
The proof of Theorem~\ref{theo:ex se} is inspired from ideas developed
in~\cite{BBD10,BBDlong}. The keypoint is to show that the existence of
a secure equilibrium in a game $(\mathcal{G},v_0)$ is equivalent to
the existence of a secure equilibrium (with two additional properties)
in the finite game $\ttree{d}$ for a well-chosen depth $d$. The
existence of the latter equilibrium is decidable. Notice that by
Corollary~\ref{coro:kuhn secure} a secure equilibrium always exists in
$\ttree{d}$; however we do not know if a secure equilibrium with the
two required additional properties always exists in $\ttree{d}$.

Let us formally introduce these two properties. The first one requires
that the secure equilibrium is \emph{\goalopt}, meaning that all the
goal sets visited along its outcome are visited for the first time
\emph{before a certain given depth}. For any game $\mathcal{G}$ played
on a graph with $|V|$ vertices by $|\Pi|$ players, we fix the
following constant: $\depthgoal:=2\cdot |\Pi| \cdot |V|$.

\begin{defi}\label{def:goalopt}
  Given a game $(\mathcal{G},v_0)$ and a strategy profile
  $(\sigma_i)_{\ipi}$ in $\mathcal{G}$, with outcome $\rho$, we say
  that $(\sigma_i)_{\ipi}$ is \emph{\goalopt} if and only if for all
  $\ipi$ such that $\cost_i(\rho) < +\infty$, we have that
  $\cost_i(\rho) < \depthgoal$.
\end{defi}

The second property asks for a secure equilibrium that is
\emph{\devopt}, meaning that whenever a player deviates from its
outcome, he realizes \emph{within a certain given number of steps}
that his deviation is not profitable for him.

\begin{defi}\label{def:devopt}
  Given a game $(\mathcal{G},v_0)$ and a secure equilibrium
  $(\sigma_i)_{\ipi}$ in $\mathcal{G}$, with outcome $\rho$, we say
  that $(\sigma_i)_{\ipi}$ is \emph{\devopt} if and only if for every
  player~$j \in \Pi$ and every strategy $\sigma'_j$ of player~$j$, we
  have that $$\cost(\rho_{< \depthdev}) \not\prec_j \cost(\rho'_{<
    \depthdev}),$$ where $\depthdev= \max\{\cost_i(\rho)~|~
  \cost_i(\rho) <+\infty\} + |V|$ and $\rho' = \langle
  \sigma'_j,\sigma_{-j} \rangle_{v_0}$.
\end{defi}
Remark that Definitions~\ref{def:goalopt} and~\ref{def:devopt} extend
to games $\ttree{d}$ where $d \ge \depthgoal$.

We can now state the key proposition for proving Theorems~\ref{theo:ex
  se} and~\ref{theo:finitemem se}. 
  
\begin{prop}\label{prop:se inf iff se fin}
  Let $(\mathcal{G},v_0)$ be a game, and $\depth=\depthgoal + 3\cdot
  |V|$.
  \begin{enumerate}[\em(1)]
  \item If there exists a \goalopt and \devopt secure equilibrium in
    \linebreak$\ttree{\depth}$, then there exists a secure equilibrium
    in $(\mathcal{G},v_0)$ that is finite-memory.
 \item If there exists a secure equilibrium in $(\mathcal{G},v_0)$,
   then there exists a \goalopt and \devopt secure equilibrium in
   $\ttree{\depth}$.
  \end{enumerate}
  \end{prop}

\noindent At this stage, it is difficult to give some intuition about the choice
of the values $\depthgoal$, $\depthdev$ and $\depth=\depthgoal +
3\cdot |V|$. These values are linked to the proofs contained in this
section.

\proof[Proof of Theorem~\ref{theo:ex se}]
%%\begin{proof}[of Theorem~\ref{theo:ex se}]
By Proposition~\ref{prop:se inf iff se fin}, there exists a secure
equilibrium in~$(\mathcal{G},v_0)$ iff there exists a \goalopt and
\devopt secure equilibrium in $\ttree{\depth}$, with
$\depth=\depthgoal + 3\cdot |V|$. The latter property is decidable in
{\sf NExpSpace} (in $|V|$ and $|\Pi|$). Indeed, $\tfin$ has an 
exponential size. Guessing a strategy profile $(\sigma_i)_{\ipi}$ in
this tree also needs an exponential size. Then we can test in
exponential size whether $(\sigma_i)_{\ipi}$ is a \goalopt and
\devopt secure equilibrium in $\ttree{\depth}$. By Savitch's theorem,
deciding the existence of a secure equilibria is thus in {\sf
  ExpSpace}.  
\qed
%%\end{proof}

\proof[Proof of Theorem~\ref{theo:finitemem se}]
%%\begin{proof}[of Theorem~\ref{theo:finitemem se}] 
This theorem is a direct consequence of Proposition~\ref{prop:se inf
  iff se fin}.  Indeed consider a secure equilibrium in a
game~$(\mathcal{G},v_0)$.  We first apply Proposition~\ref{prop:se inf
  iff se fin} (Part (\emph{ii})) to this strategy profile to get a
\goalopt and \devopt secure equilibrium $(\sigma_i)_{\ipi}$ in
$\ttree{\depth}$, for $\depth=\depthgoal + 3\cdot |V|$.  Then we apply
Proposition~\ref{prop:se inf iff se fin} (Part (\emph{i})) to the
equilibrium $(\sigma_i)_{\ipi}$, to get a finite-memory secure
equilibrium back in $(\mathcal{G},v_0)$.
%
% It remains to show that this equilibrium is a finite-memory strategy
%profile. This proof is very similar to the proof of~\cite[Proposition
%   25]{BBDlong} and thus is not given in details. Roughly speaking, a
%finite amount of memory is enough to produce the outcome
%$\alpha\beta^\omega$; outside of this outcome it is enough to
%remember how the equilibrium is defined for histories up to length
%$|\alpha|$ (after depth $|\alpha|$, memoryless strategies are used).
 \qed
 %%\end{proof}

Let us remark that in Theorem~\ref{theo:finitemem se}, the
finite-memory secure equilibrium is created from the one given by
hypothesis and the construction is made in such a way that the set of
players whose goal set is visited along the outcome is the same for
both equilibria.

The proof of Proposition~\ref{prop:se inf iff se fin} is long and
technical. The next two sections are devoted to the two parts of this
proposition.

\subsection{Part ($i$) of Proposition~\ref{prop:se inf iff se fin}}

This section is devoted to the proof of Proposition~\ref{prop:se inf
  iff se fin}, Part ($i$).  We begin with a useful characterisation of
a \devopt secure equilibrium.

\begin{lem}\label{lem:devopt}
  With the previous notations of Definition~\ref{def:devopt}, a secure
  equilibrium~$(\sigma_i)_{\ipi}$ is \devopt if and only if for every
  player~$j \in \Pi$ and every strategy $\sigma'_j$ of player~$j$, if
  \begin{enumerate}[\em(1)]
    \item $\cost_j(\rho) = \cost_j(\rho')$,
    \item $\forall \,i \in \Pi$ such that $\cost_i(\rho) < +
      \infty$, we have that $\cost_i(\rho) \le \cost_i(\rho')$,
    \item $\exists \,i \in \Pi \ \cost_i(\rho) < \cost_i(\rho')$,
  \end{enumerate}
  then there exists $l \in \Pi$ such that $\cost_l(\rho)=+\infty$ and
  $\cost_l(\rho') < \depthdev$.
\end{lem}

\proof
%%\begin{proof}
  Let us first assume that $(\sigma_i)_{\ipi}$ is a \devopt\ secure
  equilibrium whose outcome is denoted by $\rho$. Given any player~$j
  \in \Pi$, let $\sigma'_j$ be a strategy fulfilling the hypotheses of
  the lemma and $\rho'$ the outcome given by $\langle
  \sigma'_j,\sigma_{-j}\rangle_{v_0}$. Let us denote respectively by
  $(x_i)_{\ipi}$ and $(y_i)_{\ipi}$ the cost profiles of the histories
  $\rho_{< \depthdev}$ and $\rho'_{< \depthdev}$.
    
  Notice that by definition of $\depthdev$, $\cost_i(\rho) = x_i$ for
  all $i$. For $\rho'$, we have $\cost_i(\rho') = y_i$ provided
  $\cost_i(\rho') < \depthdev$.  Otherwise, it may happen that $y_i =
  +\infty$ and $\cost_i(\rho') < +\infty$. So, it holds that
  $\cost_i(\rho') \le y_i$ for all $i$. These observations will be
  often used in the sequel of the proof.
  
  Since $(\sigma_i)_{\ipi}$ is 
  \devopt, we have $\cost(\rho_{< \depthdev})
  \not\prec_j \cost(\rho'_{< \depthdev})$ meaning that:
  \begin{equation}\label{eq:nondevprecj}
    \big(x_j \le y_j\big) \wedge \big(x_j \ne y_j \vee (\exists \ipi \
    x_i > y_i ) \vee (\forall \ipi \ x_i \ge y_i)\big)\,.
  \end{equation}
  By hypothesis~$(i)$, $x_j = y_j$. By hypothesis~$(iii)$, we cannot
  have $\forall i \in \Pi ~x_i \geq y_i$. Therefore to
  satisfy~(\ref{eq:nondevprecj}), there must exist a player $i$ such
  that $x_i > y_i$.  If $\cost_i(\rho) < + \infty$, then by definition
  of $\depthdev$, $\cost_i(\rho) = x_i > y_i = \cost_i(\rho')$ in
  contradiction with hypothesis~$(ii)$.  Therefore $\cost_i(\rho) = +
  \infty$. From $x_i > y_i$, it follows that $\cost_i(\rho') <
  \depthdev$, which concludes the first implication of the proof.

  \medskip For the converse, let us now assume that
  $(\sigma_i)_{\ipi}$ is a secure equilibrium that fulfills the
  property stated in Lemma~\ref{lem:devopt}.  We will prove that it
  is \devopt, that is, for any player~$j \in \Pi$, and any deviation
  $\sigma'_j$ of player~$j$, we have that
  $\cost(\rho_{< \depthdev}) \not\prec_j \cost(\rho'_{< \depthdev})$,
  with $\rho = \langle (\sigma_i)_{\ipi}\rangle_{v_0}$ and $\rho'
  = \langle \sigma'_j,\sigma_{-j} \rangle_{v_0}$. By denoting
  respectively by $(x_i)_{\ipi}$ and $(y_i)_{\ipi}$ the cost profiles
  of $\rho_{< \depthdev}$ and $\rho'_{< \depthdev}$, it is equivalent
  to prove (\ref{eq:nondevprecj}).

  Since $(\sigma_i)_{\ipi}$ is a secure equilibrium, we know that
  $\sigma'_j$ is not a $\prec_j$-profitable deviation. In particular,
  player~$j$ can not strictly decrease his cost along $\rho'$, and
  thus $x_j \leq y_j$.  It remains to prove that the second conjunct
  of~(\ref{eq:nondevprecj}) is true. For this, we first show that as
  soon as one of the hypotheses among $(i)$, $(ii)$ or $(iii)$ is not
  fulfilled, this conjunct is satisfied.
  \begin{iteMize}{$\bullet$}
  \item If $\cost_j(\rho) < \cost_j(\rho')$, by choice of $\depthdev$,
    we also have that $x_j < y_j$. Moreover, the case $\cost_j(\rho) >
    \cost_j(\rho')$ is not possible as $(\sigma_i)_{\ipi}$ is a secure
    equilibrium.
  \item If there exists $\ipi$ such that $\cost_i(\rho) < + \infty$
    and $\cost_i(\rho) > \cost_i(\rho')$, then $x_i > y_i$.
  \item If for all $i \in \Pi$, $\cost_i(\rho) \ge \cost_i(\rho')$, we
    also have that $x_i \ge y_i$, for all $i$.
  \end{iteMize} 
  Thus the remaining deviations to consider fulfill hypotheses~$(i)$,
  $(ii)$ and $(iii)$. In this case, there exists $l \in \Pi$ such that
  $\cost_l(\rho)=+\infty$ and $\cost_l(\rho') < \depthdev$. In
  particular we have that $x_l > y_l$, and the second conjunct of
  (\ref{eq:nondevprecj}) is true.  
  \qed
%%\end{proof}

%We prove the two implications of Proposition~\ref{prop:se inf iff se
%  fin} separately, and begin with this one: given a game $\mathcal{G}$, if
%there exists a \goalopt and \devopt secure equilibrium in
%$\ttree{\depth}$, for $\depth = \depthgoal + 3\cdot |V|$, then there
%exists a secure equilibrium in~$\mathcal{G}$.

The ideas of the proof for Part ($i$) of Proposition~\ref{prop:se inf
  iff se fin} are as follows. Suppose that there exists a \goalopt and
\devopt secure equilibrium $(\sigma_i)_{\ipi}$ in $\ttree{\depth}$,
for $\depth=\depthgoal + 3\cdot |V|$.  To get from $(\sigma_i)_{\ipi}$
a finite-memory secure equilibrium in $(\mathcal{G},v_0)$, we use a
similar construction as \cite[Proposition 25]{BBDlong} where it is
shown, in the context of two-player games, how to extend a secure
equilibrium in a finite truncation of $(\mathcal{G},v_0)$ to a secure
equilibrium in $(\mathcal{G},v_0)$. The rough idea is as follows. Due
to the hypotheses, the outcome $\pi$ of $(\sigma_i)_{\ipi}$ has a
prefix $\alpha\beta$ such that all goal sets visited by $\pi$ are
already visited by $\alpha$, and such that $\beta$ is a cycle. The
required secure equilibrium is specified such that its outcome is
equal to $\alpha\beta^{\omega}$ and any deviating player is punished
by the coalition of the other players in a way that this deviation is
not profitable for him.  This secure equilibrium can be constructed in
a way to be finite-memory.  

\proof[Proof of Proposition~\ref{prop:se inf iff se fin}, Part ($i$)]
%%\begin{proof}[of Proposition~\ref{prop:se inf iff se fin}, Part ($i$)]
Let us set $\Pi=\{1,\ldots,n\}$.  Let $(\tau_i)_{\ipi}$ be a \goalopt
  and \devopt secure equilibrium in the game $\ttree{\depth}$ and
  $\pi$ its outcome.  Since $|\pi|= \depthgoal + 3\cdot |V|$, we can
  write \begin{align*} \pi = \alpha\beta\gamma \text{\ \ \ with\ \ \ }
  & \beta \mbox{ non-empty}\notag\\
  & \last(\alpha)= \last(\alpha\beta)\\ & |\alpha| \geq \depthgoal +
  |V|\notag\\ & |\alpha\beta| \leq \depthgoal + 2\cdot
  |V|\,.  \end{align*} We have that
  $\visit(\alpha)=\visit(\alpha\beta\gamma)$ (no new goal set is
  visited after $\alpha$) because $|\alpha| \geq \depthgoal$ and
  $(\tau_i)_{\ipi}$ is \goalopt. This enables us to
  use~\cite[Lemma~15]{BBDlong} as follows. Let $j \in \Pi$ be such
  that $\alpha$ does not visit $\F_j$, and suppose that player $j$
  deviates from the history $\alpha$. This lemma states that for all
  histories $hv$ consistent with $\tau_{-j}$ and such that $|hv| \le
  |\alpha\beta|$, then the coalition formed by all the players
  $i \in \Pi \setminus \{j\}$ can play to prevent player~$j$ from
  reaching his goal set $\F_j$ from vertex $v$. It means that this
  coalition has a \memoryless\ winning strategy $\strat^v_{-j}$ from
  vertex $v$ in the zero-sum reachability game $\mathcal{G}_j =
  (G_j,\F_j,V)$ (see Corollary~\ref{coro:gamei}). For each
  player~$i \not= j$, let $\strat^v_{i,j}$ be the \memoryless\
  strategy of player~$i$ in $\mathcal{G}$ induced by $\strat^v_{-j}$.

 We define a finite-memory secure equilibrium in the game $\ginf$ using
 the same idea as in the proof of \cite[Proposition 25]{BBDlong}.  The
 idea is to specify the required secure equilibrium as follows: each
 player~$i$ plays according to $\abo$ (which is the outcome of this
 equilibrium) and punishes player~$j \not= i$ if he deviates from
 $\abo$, by playing according to $\tau_i$ until depth $|\alpha|$, and
 after that, by playing arbitrarily if $\alpha$ visits $\F_j$, and
 according to $\strat^v_{i,j}$ otherwise (where $v$ is the vertex
 visited at depth $|\alpha|$ when deviating).

 Formally we first need to specify a punishment function $\pun$.  For
 the initial vertex $v_0$, we define $\pun(v_0) = \bot$ and for all
 histories $hv \in H$ such that $h \in H_i$, we let:
  \[ \pun(hv) := \left\{
  \begin{array}{ll}
    \bot & \mbox{ if $\pun(h)= \bot$ and $hv < \abo$,}\\
    i & \mbox{ if $\pun(h)= \bot$ and $hv \not< \abo$,}\\
    \pun(h) & \mbox{ otherwise ($\pun(h) \not=\bot$).} 
  \end{array}\right. \]
Then the definition of the secure equilibrium $(\sigma_i)_{\ipi}$ in
$\ginf$ is as follows. For all $\ipi$ and $h \in H_i$,
  \[ \sigma_i(h) := \left\{
  \begin{array}{ll}
    v
    & \mbox{ if $\pun(h)=\bot$ ($h<\abo$); such that
     $hv<\abo$,}\\
    \textit{arbitrary} & \mbox{ if $\pun(h)=i$,}\\
    \tau_i(h) & \mbox{ if $\pun(h) \not= \bot,i$ and
     $|h| < |\alpha|$,}\\
    \strat^v_{i,\pun(h)}(h) & \mbox{ if $\pun(h) \not= \bot,i$, 
     $|h| \ge |\alpha|$, $\alpha$ does not visit $\F_{\pun(h)}$;}\\
    & \mbox{ such that $\exists\, h'v \le h$ with $|h'v|=|\alpha|$,}\\
    \textit{arbitrary} & \mbox{ otherwise ($\pun(h) \not= \bot,i$,
     $|h| \ge |\alpha|$ and $\alpha$ visits $\F_{\pun(h)}$)},\\
  \end{array}\right.\]
  where \textit{arbitrary} means that the next vertex is chosen
  arbi\-tra\-rily (in a \memoryless\ way). Clearly the outcome
  of $(\sigma_i)_{\ipi}$ is the play $\abo$.

  Let us show that $(\sigma_i)_{\ipi}$ is a secure equilibrium
  in the game $\ginf$.  Assume by contradiction that there exists a
  $\prec_j$-profitable deviation $\sigma_j'$ for player~$j$
  w.r.t. $(\sigma_i)_{\ipi}$ in $\ginf$. 
  Let $\tau_j'$ be the strategy $\sigma_j'$ restricted to
  $\ttree{\depth}$. We are going to show that $\tau_j'$ is a
  $\prec_j$-profitable deviation  for player~$j$
  w.r.t. $(\tau_i)_{\ipi}$ in $\ttree{\depth}$, which is impossible by
  hypothesis. Here are some useful notations:
  \[
  \begin{array}{lllll}
    \pi&=&\langle (\tau_i)_{\ipi} \rangle_{v_0}
    &\text{ of cost profile } & (x_1,\ldots,x_n)\\
    \pi'& =&\langle \tau_j', \tau_{-j} \rangle_{v_0}
    &\text{ of cost profile } & (x_1',\ldots,x_n')\\
    \rho &=&\langle (\sigma_i)_{\ipi} \rangle_{v_0}
    &\text{ of cost profile } & (y_1,\ldots,y_n)\\
    \rho'&=&\langle \sigma_j',\sigma_{-j} \rangle_{v_0}
    &\text{ of cost profile } & (y_1',\ldots,y_n').
  \end{array} \]
  Notice that the play $\pi'$ coincide with the play $\rho'$ at least
  until depth $|\alpha|$ (by definition of $\tau_j'$ and
  $\sigma_{-j}$); they can differ afterwards.  Clearly $\pi$ and $\rho$ 
  coincide at least until depth $|\alpha\beta|$. The situation
  is depicted in Fig.~\ref{fig:ES plays}.

\begin{figure}[h!]
  \centering
  \begin{tikzpicture}[yscale=.6,xscale=.8]
    \everymath{\scriptstyle}
    \draw (0,10) -- (-5,0);
    \draw (0,10) -- (5,0);
    
    \draw[very thin, dashed] (-3,6) -- (3,6);
    \draw (3.1,6) node[right] (q0) {$|\alpha|$};

    \draw (.4,7.5) node[right] (q0) {$\alpha$};

    \draw (-3.5,3) -- (3.5,3);
    \draw (3.6,3) node[right] (q0) {$\depth$};

    \draw[fill=black] (0,10) circle (2pt);
    \draw[] (0,10) .. controls (.2,8) .. (1,6);
    \draw (.8,5.5) node[left] (q0) {$\beta$};
    \draw (.6,4.5) node[left] (q0) {$\beta$};
    \draw (.4,3.5) node[left] (q0) {$\beta$};
    \draw (.2,2.5) node[left] (q0) {$\beta$};
    \draw (0,1.5) node[left] (q0) {$\beta$};
    \draw (-.2,.5) node[left] (q0) {$\beta$};

    \draw[fill=black] (1,6) circle (2pt); 

    \draw[] (.8,5) .. controls (1.7,4) and (2.1,3.5) .. (2.2,3);
    \draw (2.2,3) node[below] (q0) {$\pi$};
    \draw (2.2,2.7) node[below] (q0) {$(x_i)_{\ipi}$};

    \draw[fill=black] (-1.72,5) circle (2pt);
    \draw[] (-1.72,5) .. controls (-1,4) and (-.9,3.5) .. (-.9,3);
    \draw (-.9,3.1) node[below] (q0) {$\pi'$};
    \draw (-.96,2.5) node[below] (q0) {$(x_i')_{\ipi}$};

 %   \draw (1.9,4.9) node[below] (q0) {$\tilde{\rho}$};  
   
    \draw[fill=black] (1,6) circle (2pt);
    \draw[thick] (1,6) .. controls (.5,5.5) and (.8,5) .. (.8,5);
    
    \begin{scope}[yshift=-1cm,xshift=.8cm]
      \draw[fill=black] (0,6) circle (2pt);
      \draw[thick] (0,6) .. controls (-.5,5.5) and (-.2,5) .. (-.2,5);
    \end{scope}
    \begin{scope}[yshift=-2cm,xshift=.6cm]
      \draw[fill=black] (0,6) circle (2pt);
      \draw[thick] (0,6) .. controls (-.5,5.5) and (-.2,5) .. (-.2,5);
    \end{scope}
    \begin{scope}[yshift=-3cm,xshift=.4cm]
      \draw[fill=black] (0,6) circle (2pt);
      \draw[thick] (0,6) .. controls (-.5,5.5) and (-.2,5) .. (-.2,5);
    \end{scope}
    \begin{scope}[yshift=-4cm,xshift=.2cm]
      \draw[fill=black] (0,6) circle (2pt);
      \draw[thick] (0,6) .. controls (-.5,5.5) and (-.2,5) .. (-.2,5);
    \end{scope}
    \begin{scope}[yshift=-5cm]
      \draw[fill=black] (0,6) circle (2pt);
      \draw[thick] (0,6) .. controls (-.5,5.5) and (-.2,5) .. (-.2,5);
    \end{scope}
    \begin{scope}[yshift=-6cm,xshift=-.2cm]
      \draw[fill=black] (0,6) circle (2pt);
    \end{scope}

    \draw (-.2,-.15) node[below] (q0) {$\rho$};
    \draw (-.2,-.5) node[below] (q0) {$(y_i)_{\ipi}$};

    \draw[fill=black] (.18,8.5) circle (2pt);
    \draw[] (.18,8.5) .. controls (-3,5) and (-1,2) .. (-4.5,0);
    \draw (-4.5,0) node[below] (q0) {$\rho'$};
    \draw (-4.56,-.5) node[below] (q0) {$(y_i')_{\ipi}$};
  \end{tikzpicture}
  \caption{Plays~$\pi$ and~$\rho$, and their respective deviations~$\pi'$
    and~$\rho'$.}
  \label{fig:ES plays}
\end{figure}

  \noindent As $\sigma_j'$ is a $\prec_j$-profitable deviation for player~$j$
  w.r.t. $(\sigma_i)_{\ipi}$, we have that 
  \begin{eqnarray}\label{eq:devprof}
  (y_1,\ldots,y_n) \prec_j (y_1',\ldots,y_n').
  \end{eqnarray}
  Let us show that $\tau_j'$ is a
  $\prec_j$-profitable deviation for player~$j$
  w.r.t. $(\tau_i)_{\ipi}$, i.e.,
  $$(x_1,\ldots,x_n) \prec_j (x_1',\ldots,x_n').$$ 
  By~(\ref{eq:devprof}), one of the next three cases stands.
  \begin{enumerate}[(1)]
  \item[(1)] $y_j' < y_j < +\infty$. 

    \smallskip
    As $\rho=\abo$ and $\visit(\alpha)=\visit(\alpha\beta\gamma)$, 
    it means that $\alpha$ visits $\F_j$, and then $y_j = x_j$.
    Since $y_j'<|\alpha|$, we also have
    $x_j'=y_j'$ (as $\pi'$ and $\rho'$ coincide until depth
    $|\alpha|$). Therefore $x_j' < x_j$, and $(x_1,\ldots,x_n) \prec_j
    (x_1',\ldots,x_n')$.
    \smallskip
  \item[(2)] $y_j' < y_j = +\infty$. 
    
    \smallskip
    If $y_j' \leq |\alpha|$, we have again $x_j'=y_j'$.  
    Since $\visit(\alpha)=\visit(\pi)$, it follows that
    $x_j=y_j=+\infty$. Thus $x_j' < x_j$, and so $(x_1,\ldots,x_n) \prec_j
    (x_1',\ldots,x_n')$.

    We show that the case $y_j' > |\alpha|$ is impossible.  By
    definition of $\sigma_{-j}$, the play $\rho'$ is consistent with
    $\tau_{-j}$ until depth $|\alpha|$, and then with $\strat^v_{-j}$
    from $\rho'_{|\alpha|}$ (as $y_j=+\infty$). The play $\rho'$ cannot
    visit $\F_j$ after a depth $> |\alpha|$ by definition of
    $\strat^v_{-j}$.
    \smallskip
  \item[(3)] $y_j=y_j'$, $\forall \ipi\ y_i \le y_i'$ and $\exists\, \ipi\
    y_i < y_i'$.
    
    \smallskip
    The fact that $y_j=y_j'$ implies $y_j=x_j \ge x_j'$ (as
    $\visit(\alpha)=\visit(\pi)$). If $x_j' < x_j$, then
    $(x_1,\ldots,x_n) \prec_j (x_1',\ldots,x_n')$. 
    
    We show that the case
    $x_j'=x_j $ is impossible. We can show that for all $i \in \Pi$ such that
    $x_i < +\infty$, we have $x_i \le x_i'$, and that  there exists $\ipi$
    such that $x_i < x_i'$. Since $(\tau_i)_{\ipi}$ is \devopt,
    Lemma~\ref{lem:devopt} implies that there exists some $l \in \Pi$
    such that $x_l=+\infty$, and $x'_l < \depthdev =
    \max\{x_i~|~ x_i <+\infty\} + |V|$. As
    $(\tau_i)_{\ipi}$ is also \goalopt, we have that $\depthdev \le
    \depthgoal + |V| \leq |\alpha|$. As $\rho'$ is consistent with
    $\tau_{-j}$ until depth $|\alpha|$, it follows that $y_l' = x'_l<
    y_l = x_l = +\infty$. Thus case (3) is impossible.
  \end{enumerate}
  
  Therefore, each case is either impossible or shows that
  $(x_i)_{\ipi} \prec_j (x_i')_{\ipi}$. This is in contradiction with
  $(\tau_i)_{\ipi}$ being a secure equilibrium in $\ttree{\depth}$,
  and therefore, $(\sigma_i)_{\ipi}$ is a secure equilibrium in
  $\ginf$, thus in $(\mathcal{G},v_0)$.
    
  It remains to show that $(\sigma_i)_{\ipi}$ is a finite-memory
  strategy profile. This proof is very similar to the proof
  of~\cite[Proposition 25]{BBDlong} and thus is not given in
  details. Roughly speaking, a finite amount of memory is enough to
  produce the outcome $\alpha\beta^\omega$; outside of this outcome it
  is enough to remember how $(\sigma_i)_{\ipi}$ is defined for
  histories up to length $|\alpha|$ (after depth $|\alpha|$,
  memoryless strategies are used).  
  \qed
%%\end{proof}

\begin{rem}\label{rem:(i) costs}
  This proof shows in fact a little stronger result: if there exists a
  \goalopt and \devopt secure equilibrium in $\ttree{\depth}$, then
  there exists a finite-memory secure equilibrium in $(\mathcal{G},v_0)$
  \emph{with the same cost profile}.
\end{rem}

\subsection{Part ($ii$) of Proposition~\ref{prop:se inf iff se fin}}

Part ($ii$) of Proposition~\ref{prop:se inf iff se fin} states that if
there exists a secure equilibrium in a game $(\mathcal{G},v_0)$, then
there exists a \goalopt and \devopt secure equilibrium in
$\ttree{\depth}$, for $\depth=\depthgoal + 3\cdot |V|$.  The proof
needs several steps. Suppose that there exists a secure equilibrium
$(\sigma_i)_{\ipi}$ in $(\mathcal{G},v_0)$. The first step consists in
transforming $(\sigma_i)_{\ipi}$ into a \goalopt and \devopt secure
equilibrium in $(\mathcal{G},v_0)$ (Proposition~\ref{prop:goalopt
  devopt}); the second step in showing that its restriction to $\tfin$
with $\depth=\depthgoal + 3\cdot |V|$ is still a \goalopt and \devopt
secure equilibrium in $\ttree{d}$.

\begin{prop}\label{prop:goalopt devopt}
If there exists a secure equilibrium in a game $(\mathcal{G},v_0)$,
then there exists one in $(\mathcal{G},v_0)$ which is \goalopt and
\devopt.
\end{prop}

To get a \goalopt equilibrium, the idea is to eliminate some
unnecessary cycles (see Definition~\ref{def:unnecessary cycle}). Such
an idea has already been developed in \cite[Lemma 19]{BBDlong} for
Nash equilibria. Unfortunately, this lemma cannot be applied for
secure equilibria (as shown in Example~\ref{ex:se}). Adapting it to
the context of secure equilibria is not trivial, the underlying
constructions are more involved: we need to modify the strategies of
the coalition against a deviating player. By using specific punishing
strategies for the coalitions, we are then able to get a
\goalopt equilibrium that is also \devopt, due to the particular form
of these strategies.

\begin{exa}\label{ex:se}
  Consider the three-player game of Fig.~\ref{fig:secycle} initialized
  at $A$, where $V_1 = \{A,C,D\}$, $V_2 = \{B\}$ and $V_3=\emptyset$,
  $\F_1=\F_2=\{A\}$ and $\F_3=\{D\}$. The strategy profile
  $(\sigma_1,\sigma_2,\sigma_3)$ defined\footnote{The stategy
    $\sigma_3$ of player~$3$ has not to be defined as
    $V_3=\emptyset$.} below is a secure equilibrium whose outcome is
  $ABCBD^\omega$ and cost profile $(0,0,5)$:
$$\sigma_1(h)= 
 \begin{cases}
   B & \text{ if } h=A \text{ or } ABC,\\
   D & \text{ otherwise.}
   \end{cases} 
   \quad ; \quad \sigma_2(h)=
 \begin{cases}
        C & \text{ if } h=AB, \\
        D & \text{ otherwise.}
      \end{cases}
$$ In Example~\ref{ex:cycle}, we gave two equilibria whose outcome has
 unnecessary cycles.  Here, we also face such a situation, with the
 cycle $BCB$. If we modify $(\sigma_1,\sigma_2,\sigma_3)$ in order to
 remove this cycle, as done in \cite[Lemma 19]{BBDlong} for Nash
 equilibria, the resulting strategy profile is a Nash equilibrium with
 outcome $ABD^\omega$ and cost profile $(0,0,3)$, however it is no
 longer a secure equilibrium. Indeed player~$1$ has a
 $\prec_1$-profitable deviation by taking the edge $(A,D)$ instead of
 $(A,B)$, which leads to a cost of $4$ for player~$3$ (instead of
 $3$). In the sequel we show how to modify the approach of~\cite[Lemma
   19]{BBDlong} in a way to keep the property of secure equilibrium.

\begin{figure}[h!]
    \centering
    \begin{tikzpicture}[->,node distance=1.6cm,bend angle=20] 
      \everymath{\scriptstyle} \tikzstyle{j0}=[draw,circle,text
        centered,,inner sep=2pt] \tikzstyle{j1}=[draw,rectangle,text
        centered,,inner sep=4pt]
      \tikzstyle{j2}=[draw,double,circle,fill=black!20!white,text
        centered,inner sep=2pt] \tikzstyle{j3}=[draw,circle,very
        thick,text centered,inner sep=2pt]
      
      \node[j2]  (0) {$A$};
      \node[j1]  (1) [right of=0] {$B$};
      \node[j0]  (2) [right of=1] {$C$};
      \node[j3]  (3) [below of=1] {$D$};
      \node (4) [left=0,xshift=-0.5cm] {};
      \path
      (4) edge (0)
      (0) edge (1)
      (1) edge[bend right]  (2)
      (2) edge[bend right]  (1)
      (0) edge node[right]{$4$}  (3)
      (1) edge node[right]{$2$}  (3)
      (2) edge  (3)
      (3) edge[loop below] (3);

    \end{tikzpicture} 
    \caption{A three-player game with $\F_1=\F_2=\{A\}$ and
      $\F_3=\{D\}$.}
\label{fig:secycle} 
\end{figure}
\end{exa}

In order to prove Proposition~\ref{prop:goalopt devopt}, we need three
lemmas: Lemmas~\ref{lem:tec},~\ref{lem:remove a cycle}
and~\ref{lem:es devopt}.  Given a secure equilibrium,
Lemma~\ref{lem:tec} describes some particular memoryless strategies
for the coalition when a player deviates. Lemma~\ref{lem:remove a
  cycle} (counterpart of \cite[Lemma 19]{BBDlong} for secure
equilibria) states that we can remove a cycle from the outcome of a
secure equilibrium, but the strategies have to be somewhat modified
with these specific coalition strategies. This lemma is used in the
proof of Proposition~\ref{prop:goalopt devopt} to get a
\goalopt secure equilibrium. Lemma~\ref{lem:es devopt} states that we
can also get a \devopt secure equilibrium.
% and its proof is based on
%Lemma~\ref{lem:tec}. Then, Proposition~\ref{prop:goalopt devopt} and
%Part $(ii)$ of Proposition~\ref{prop:se inf iff se fin} follow.

%In order to demonstrate Proposition~\ref{prop:goalopt devopt} and
%finish the proof of Proposition~\ref{prop:se inf iff se fin}, we need
%some preliminary results. First, we give some particular memoryless
%strategies for the coalition when a player deviates from the
%equilibrium, and then, we show how to remove a cycle from its outcome.

\subsubsection*{Memoryless coalition strategies.}

\quad Given a secure equilibrium in a game $(\mathcal{G},v_0)$, we
here prove the existence of interesting memoryless strategies for the
coalition against a deviating player.

Let us first introduce the definition of a
\emph{$j$-\promising history} for some deviating
player~$j$. Intuitively player~$j$ deviates from a strategy profile
$(\sigma_i)_{\ipi}$ and constructs a history $h$ consistent with
$\sigma_{-j}$.  This history $h$ is called
$j$-\promising w.r.t. $(\sigma_i)_{\ipi}$ if player~$j$ does not know
yet if this deviation will be \emph{$\prec_j$-profitable} for him
w.r.t. $(\sigma_i)_{\ipi}$, but he can still \emph{hope} that it will
be, without knowing what he will play after $h$.

\begin{defi}\label{def:promising}
  Let $(\sigma_i)_{\ipi}$ be a strategy profile in a game
  $(\mathcal{G},v_0)$, with cost profile $(x_i)_{\ipi}$. Let us assume
  that $\Pi=\{1,\ldots,n\}$ and
  $$x_1 \le \ldots \le x_k < x_{k+1} \le \ldots \le x_{n}$$ where $0
  \le k < n$.  Let $h$ be a history of the game such that $x_k \le |h|
  < x_{k+1}$.

  For any player~$j \in \Pi$, we say that $h$ is
  \emph{$j$-\promising}\ w.r.t. $(\sigma_i)_{\ipi}$ if $h$ is consistent
  with $\sigma_{-j}$ and if
  \begin{iteMize}{$\bullet$}
  \item in the case where $x_{k+1} < +\infty$: 
    \begin{iteMize}{$-$}
    \item if $j \le k$, we have that $\cost_j(h) = x_j$ and
      $\forall\, \ipi$  $\cost_i(h) \ge x_i$,
    \item if $j > k$, we have that
      $\cost_j(h) = +\infty$;
  \end{iteMize}
  \item in the case where $x_{k+1}=+\infty$: \\
    $\cost_j(h) = x_j$,  $\forall\, \ipi$  $\cost_i(h) \ge x_i$ and
    $\exists\, \ipi$ $\cost_i(h) > x_i$.
  \end{iteMize}
\end{defi}
 
\noindent In the case where $x_{k+1} < +\infty$ and $j \le k$, \emph{along $h$},
player $j$ has been able to get the same cost as along $\rho$
($\cost_j(h) = x_j$) and to not decrease the cost of the other players
($\cost_i(h) \ge x_i$). \emph{After $h$}, he hopes to be able to play
such that the resulting deviation $h\rho'$ will satisfy $(x_i)_{\ipi}
\prec_j \cost(h\rho')$.  In the case where $j > k$, player $j$ has not
visited his goal set along $h$, so he does not know yet if his
deviation will be $\prec_j$-profitable for him. However he hopes to
visit it early enough after $h$ along $h\rho'$, such that
$\cost_j(h\rho') < x_j$, or to get the same cost while increasing the
cost of the other players in a way that $(x_i)_{\ipi} \prec_j
\cost(h\rho')$. 

In the case where $x_{k+1}=+\infty$, the history $\rho_{\leq |h|}$ has
visited all the goal sets $\F_i$ such that $\cost_i(\rho) < +
\infty$. Thus player~$j$ could have a $\prec_j$-profitable deviation
$h\rho'$ if he can avoid visiting the goal sets $\F_i$, where
$i \ge k+1$ ($i \not= j$).

%\begin{definition}\label{def:promising}
 % Let $\game$ be a game, $(\sigma_i)_{\ipi}$ be a strategy profile
%  \fbox{juste un profil de strategie c'est suffisant non? pas oblige
 %   secure dans cette def?}, $\rho=\langle (\sigma_i)_{\ipi} \rangle$
 % be its outcome and $(x_i)_{\ipi}$ be its cost profile. Let us assume
  %that $\Pi=\{1,\ldots,n\}$ and
 % $$x_1 \le \ldots \le x_k < x_{k+1} \le \ldots \le x_l < x_{l+1} =
  %\ldots = x_{n} = +\infty$$ where $0 \le k \le \fbox{besoin de $<$??} l
  %\le n$.  Let $h$ be a history of the game such that $x_k \le |h| <
  %x_{k+1}$.
%
  %For any player~$j \in \Pi$, we say that $h$ is
  %\emph{$j$-\promising}\ w.r.t. $(\sigma_i)_{\ipi}$ if $h$ is consistent
 % with $\sigma_{-j}$ and if
 % \begin{iteMize}
  %\item in the case $j \le k$, we have that $\cost_j(h) = x_j$ and
  %  $\forall \ipi$,  $\cost_i(h) \ge x_i$;
  %\item in the case $j > k$, we have that
  %  $\cost_j(h) = +\infty$.
  %\end{iteMize}
%\end{definition}

Given a $j$-\promising history $h$ of player $j$, the next lemma
describes the existence of interesting memoryless strategies of the
coalition $\Pi \setminus \{j\}$ from the last vertex of $h$. This
lemma uses some qualitative two-player zero-sum reachability under
safety games $\mathcal{G}_{-j}=(G_{-j},R,S)$ associated with the
coalition $\Pi \setminus \{j\}$ (where $G_{-j}=(V,V\setminus
V_j,V_j,E)$). In such games, the coalition $\Pi \setminus \{j\}$ aims
at reaching $R$ while staying in $S$, and player~$j$ wants to prevent
this.

\begin{lem}\label{lem:tec}
  Let $(\sigma_i)_{\ipi}$ be a secure equilibrium in a game
  $(\mathcal{G},v_0)$, with outcome $\rho$ and cost profile
  $(x_i)_{\ipi}$. Let $h$ be a $j$-\promising history
  w.r.t. $(\sigma_i)_{\ipi}$ for some player~$j \in \Pi$. Let us
  assume w.l.o.g. that $\Pi=\{1,\ldots,n\}$. If
  $$x_1 \le \ldots \le x_k \le |h| < |h| + |V| \le x_{k+1} \le \ldots
  \le x_l < x_{l+1} = \ldots = x_{n} = +\infty,$$ where $0 \le k \le l
  \le n$, then the coalition~$\Pi \setminus \{j\}$ has a memoryless
  winning strategy $\stropt_{-j}^v$ from $v=\last(h)$ in the
  qualitative two-player zero-sum game ${\mathcal{G}_{-j} =
    (G_{-j},R,S)}$ where
\begin{iteMize}{$\bullet$}
\item if $j \le k$, then $R=\displaystyle\cup_{i>k} \F_i$, and $S=V$,
\item if $k <j \le l$, then $R=V$,  and $S=V\setminus \F_j$,
\item if $l <j$ and $\cost(\rho_{\le |h|}) \preceq_j \cost(h)$, then
  $R=\displaystyle\cup_{\underset{i \not=j}{i>k}} \F_i $, and
  $S=V\setminus \F_j$,
\item if $l <j$ and $\cost(\rho_{\le |h|}) \not\preceq_j \cost(h) $,
  then $R=V$, and $S=V\setminus \F_j$.
\end{iteMize}
\end{lem}
In this lemma, either all goal sets are visited by $\rho$ and $l=n$,
or $l<n$ and the last visited goal set is $\F_l$.
Also notice that $R \not= \emptyset$ in all cases. Indeed, $k
\not= n$ as $h$ is $j$-\promising, and then the set $R$ in the case $j
\le k$ of this lemma is not empty. In the third case, it is not empty
either, otherwise we would have $k+1 = l+1 =n = j$ but such a
situation is impossible because $h$ is
$j$-\promising w.r.t. $(\sigma_i)_{\ipi}$ (see the last case of
Definition~\ref{def:promising}) and $(\sigma_i)_{\ipi}$ is a secure
equilibrium .

\proof[Proof of Lemma~\ref{lem:tec}]
%%\begin{proof}[of Lemma~\ref{lem:tec}]
By contradiction assume that the coalition~$\Pi \setminus \{j\}$ has
no winning strategy from $v$ in the game ${\mathcal{G}_{-j} =
  (G_{-j},R,S)}$, i.e.  no winning strategy from $v$ to reach $R$ while
staying in $S$.  By Corollary~\ref{coro:gamei}, it implies that
player~$j$ has a memoryless winning strategy $\stropt_j^v$ from $v$ to
stay outside $R$ or to reach $V \setminus S$. Recall that $h$ is
consistent with $\sigma_{-j}$ as it is
$j$-\promising w.r.t. $(\sigma_i)_{\ipi}$.  Let $\rho'$ be the play
with prefix $h$ that is consistent with $\sigma_{-j}$, and with
$\stropt_j^v$ from $v$ (see Fig.~\ref{fig:coal}).  In the four cases
of the lemma, we then prove that $(x_i)_{\ipi} \prec_j
(\cost_i(\rho'))_{\ipi}$, meaning that player~$j$ has a
$\prec_j$-profitable deviation w.r.t. $(\sigma_i)_{\ipi}$, which is
impossible.

\begin{figure}[h!]
    \centering
    \begin{tikzpicture}[yscale=.5,xscale=.8]
      \everymath{\scriptstyle}
      \draw (0,8) -- (-5,-2);
      \draw (0,8) -- (5,-2);

%      \draw[fill=black] (0,8) circle (1pt);

      \draw[very thin,dotted] (-4,0.5) -- (5.1,0.5);
      \draw[very thin,dotted] (-4,3) -- (5.1,3);
      \draw[very thin,dotted] (-4,1.5) -- (5.1,1.5);
      \draw[very thin,dotted] (-4,4) -- (5.1,4);

       \path (5.1,4) node[right] (q0) {$x_k$}; 
       \path (5.1,3) node[right] (q0) {$|h|$}; 
       \path (5.1,1.5) node[right] (q0) {$|h|+|V|$}; 
       \path (5.1,0.5) node[right] (q0) {$x_{k+1}$}; 

      \draw[thick] (0,8) .. controls (-.2,7.1) .. (.1,6.4);
      \draw[fill=black] (.1,6.4) circle (2pt);
      \path (.05,6.3) node[right] (q0) {$\F_1$};

%      \path (.1,7) node[right] (q0) {$h$};

      \draw[thick] (.1,6.4) .. controls (.7,3.6) .. (.8,-2);
      \path (.8,-2) node[below] (q0) {$\rho$};

      \draw[fill=black] (.77,0.5) circle (2pt);
      \path (.77,0.5) node[right] (q0) {$\F_{k+1}$};
      \draw[fill=black] (.8,-1.3) circle (2pt);
      \path (.8,-1.3) node[right] (q0) {$\F_l$};

      \draw[thick] (.3,5.5) .. controls (-1,4) .. (-1.2,3);
      \draw[fill=black] (-1.2,3) circle (2pt);

      \path (-0.1,5.1) node[left] (q0) {$h$};
      \path (-1.2,3) node[right] (q0) {$v$};

      \draw[thick] (-1.2,3) .. controls (-1.4,2) .. (-2.7,-2);
      \path (-2.7,-1.9) node[below] (q0) {$\rho'$};

      \draw[fill=black] (.3,5.5) circle (1pt);

      \draw[fill=black] (.57,4) circle (2pt);
      \path (.57,4) node[right] (q0) {$\F_k$};

    \end{tikzpicture}
    \caption{Play $\rho$ and its deviation $\rho'$ with prefix $h$.}
    \label{fig:coal}
  \end{figure}

\begin{iteMize}{$\bullet$}
\item $j \le k$.
  
  \smallskip
  The strategy $\stropt_j^v$ enables to avoid all goal sets $\F_i$
  where $i>k$. As $h$ is $j$-\promising, we have that 
  $\cost_j(h) = x_j$ and $\forall \ipi$,  $\cost_i(h) \ge x_i$. By 
  construction of $\rho'$ and as  $x_k \le |h| < x_{k+1}$, we have that
  \[\begin{array}{l}
    \cost_j(\rho') = \cost_j(h)=x_j,\\ 
    \forall i \le k,\ \cost_i(\rho') \ge x_i,\\ 
    \forall i > k,\ \cost_i(\rho') = +\infty.
  \end{array}\]
  Then for all $\ipi$, we have that $\cost_i(\rho') \ge x_i$. It
  remains to show that the cost of one player is strictly increased in
  $\rho'$ compared with $\rho$. In the case where $x_{k+1} < +\infty$,
  i.e. $k < l$, we have in particular that $x_l < +\infty$ and
  $\cost_l(\rho') = + \infty$. And in the case where $x_{k+1} =
  +\infty$ ($k = l$), we have that $(x_i)_{\ipi} \prec_j \cost(h)$ (by
  definition of $j$-\promising), i.e. there exists $\ipi$ such that
  $x_i < \cost_i(h)$. Either $\cost_i(h) = \cost_i(\rho')$ and then
  $x_i < \cost_i(\rho')$, or $\cost_i(h) = +\infty > \cost_i(\rho')$
  and so $x_i \le |h| < \cost_i(\rho')$. In both cases, it implies
  that $(x_i)_{\ipi} \prec_j (\cost_i(\rho'))_{\ipi}$.
  \smallskip
\item $k <j \le l$.
  
  \smallskip
  As $\stropt_j^v$ is memoryless, this strategy enables player~$j$ to
  reach his goal set $\F_j$ from $v$ within $|V|$ steps. Thus, we have
  that $$\cost_j(\rho') < |h|+|V| \le x_{k+1} \le x_j$$ since $k <j
  \le l$, and so, $(x_i)_{\ipi} \prec_j (\cost_i(\rho'))_{\ipi}$.
  \smallskip
\item $l <j$ and $\cost(\rho_{\le |h|}) \preceq_j \cost(h) $.
  
  \smallskip
  The strategy $\stropt_j^v$ enables to avoid all goal sets $\F_i$
  where $i>k$ and $i \not= j$, or to visit the goal set $\F_j$. On one
  hand, if $\rho'$ visits $\F_j$, then $$\cost_j(\rho') < +\infty =
  x_j$$ as $j > l$, and so, $(x_i)_{\ipi} \prec_j
  (\cost_i(\rho'))_{\ipi}$. On the other hand, if $\rho'$ does not
  visit $\F_j$, then $\rho'$ does not visit either any $\F_i$ with
  $i>k$.  Since $\cost(\rho_{\le |h|}) \preceq_j \cost(h) $, the
  situation is quite similar to the first case, and we can deduce that
 \[\begin{array}{l}
    \cost_j(\rho')  =x_j = +\infty,\\ 
    \forall i \le k,\ \cost_i(\rho') \ge x_i,\\ 
    \forall i > k,\ \cost_i(\rho') = +\infty.
  \end{array}\]
  Thus, for all $\ipi$, we have that $\cost_i(\rho') \ge
  x_i$. Moreover, exactly like in the case $j \le k$, we can show that
  there exists $\ipi$ such that $x_i < \cost_i(\rho')$. Then it
  implies that $(x_i)_{\ipi} \prec_j (\cost_i(\rho'))_{\ipi}$.
\smallskip
\item $l <j$ and $\cost(\rho_{\le |h|}) \not\preceq_j \cost(h) $.
  
  \smallskip
  Like in the second case, the strategy $\stropt_j^v$ enables
  player~$j$ to reach his goal set $\F_j$ from $v$. Then we have
  that $$\cost_j(\rho') < +\infty = x_j$$ 
  and so, $(x_i)_{\ipi} \prec_j (\cost_i(\rho'))_{\ipi}$.
  \qed
\end{iteMize}
%%\end{proof}

\subsubsection*{Removing a cycle.}

\quad The next lemma states that it is possible to modify the strategy profile
of a secure equilibrium in a way to eliminate an unnecessary cycle in its
outcome. In the notations of this lemma, notice that $\beta$ is the eliminated 
cycle (condition $ \last(\alpha)=\last(\alpha\beta)$), notice also 
that a new goal set is visited after $\alpha\beta\gamma$ 
(condition $\visit(\rho) \not= \visit(\alpha)$). 
The elimination of the cycle is possible by modifying the strategies of 
the coalitions into strategies as described in Lemma~\ref{lem:tec}.

\begin{lem}\label{lem:remove a cycle}
 Let~$(\sigma_i)_{\ipi}$ be a secure equilibrium in a
 game~$(\mathcal{G},v_0)$, with outcome $\rho$. Suppose that~$\rho =
 \alpha\beta\gamma\rhobar$, with $\beta$ non-empty and $|\gamma| \ge
 |V|$, such that
  \begin{align*}
    & \visit(\alpha)=\visit(\alpha\beta\gamma)\\ 
    & \visit(\rho) \not= \visit(\alpha)\\
    & \last(\alpha)=\last(\alpha\beta).
  \end{align*}
  Then there exists a secure equilibrium $(\tau_i)_{\ipi}$ in
  $(\mathcal{G},v_0)$ with outcome $\alpha\gamma\rhobar$.
\end{lem}

\proof
%%\begin{proof}
  Let $(x_i)_{\ipi}$ be the cost profile of $\rho$.  Let us assume w.l.o.g. that
  $\Pi=\{1,\ldots,n\}$ and $$x_1 \le \ldots \le x_k \le |\alpha| <
  |\alpha\beta\gamma| \le x_{k+1} \le \ldots \le x_l < x_{l+1} =
  \ldots = x_{n} = +\infty,$$ where $0 \le k < l \le n$ (remark that
  $k<l$ as $\visit(\rho) \not = \visit(\alpha)$).

  Let us define the required strategy profile $(\tau_i)_{\ipi}$ with
  the aim to get the outcome $\alpha\gamma\rhobar$ by eliminating
  $\beta$ in $\rho$.    For all $\ipi$ and all histories $h \in H_i$,
  we set
  \[ \tau_i(h) := \left\{
  \begin{array}{ll}
    \sigma_i(\alpha\beta\delta) & \mbox{ if $h=\alpha\delta$,}\\

    \textit{arbitrary} & \mbox{ if $\pun(h)=i$,}\\

    \stropt_{i,\pun(h)}^v(h) & \mbox{ if $\alpha \not\le h$, $\pun(h)
      \not= \bot,i$ and $\exists\, h'v$ that is $\pun(h)$-\promising}\\

    & \mbox{ w.r.t. $(\sigma_i)_{\ipi}$ and verifies $h'v \le h$ and
      $|h'v|=|\alpha|$,}\\

    \sigma_i(h) &
    \mbox{ otherwise,}
  \end{array}\right.\]
  In this definition, \textit{arbitrary} means that the next vertex is
  chosen arbi\-tra\-rily, and the punishment function $\pun$ is
  defined as in the proof of Proposition~\ref{prop:se inf iff se fin},
  Part~($i$) (adapted to the play $\alpha\gamma\rhobar$).  Moreover,
  when a player $j$ deviates, each player $i\neq j$ plays according to
  $\sigma_ i$, except in the case of a $j$-\promising history~$h$ of
  length $|\alpha|$ from which he plays according to $\stropt_{-j}^v$,
  with $v=\last(h)$ (see Lemma~\ref{lem:tec}). Notation
  $\stropt_{i,j}^v$ means the memoryless strategy of player~$i$
  induced by $\stropt_{-j}^v$.
  
  We observe that the outcome of $(\tau_i)_{\ipi}$ is the play
  $\pi=\alpha\gamma\rhobar$ (see Fig.~\ref{fig:cycle}
  and~\ref{fig:remove cycle}). Let us write its cost profile as
  $(y_1,\ldots,y_n)$. It follows that for all $\ipi$, $y_i \leq
  x_i. \label{eq:y_i + pt x_i}$ More precisely,
  \begin{align}
    & \text{- if } i \le k, 
    \text{ then } y_i=x_i; \label{eq:x_i=y_i fin}\\%\notag\\
    & \text{- if } k < i \le l, 
    \text{ then } y_i=x_i-(|\beta|+1); \label{eq:x_i>y_i}\\
    & \text{- if } i > l, \text{ then } y_i=x_i=+\infty.\label{eq:x_i=y_i}
  \end{align}

  \begin{figure}[h!]
  \null\hfill
  \begin{minipage}[b]{0.48\linewidth}
    \centering
    \begin{tikzpicture}[yscale=.5,xscale=.5]
      \everymath{\scriptstyle}
      \draw (0,10) -- (-5,0);
      \draw (0,10) -- (5,0);

      \draw[black!20!white,fill=black!20!white] (.5,7) -- (4,0) -- (-3.5,0);
      \draw[black!10!white,fill=black!10!white] (-.5,4) -- (1.5,0) -- (-2.5,0);

      \draw[fill=black] (0,10) circle (2pt);
      \draw[thick] (0,10) .. controls (-.5,8.5) .. (.5,7);

      \draw[fill=black] (.5,7) circle (2pt);
      \draw[] (.5,7) .. controls (1.2,5) and (-.2,5) .. (-.5,4);

      \draw[fill=black] (-.5,4) circle (2pt);
      \draw[thick] (-.5,4) .. controls (-1,3) and (-.5,2) .. (-1.5,0);

      %draw[fill=black] (-.2,8.1) circle (2pt);
      %draw[thick,dashed] (-.2,8.1) .. controls (-3,5) and (-1,2) .. (-4.5,0);

      \draw[fill=black] (-.825,2.3) circle (2pt);
      \draw[thick,dashed] (-.825,2.3) .. controls (1,.8)  .. (1.2,0);

      \draw (-1.5,-.1) node[below] (q0) {$\rho = \langle
        (\sigma_i)_{\ipi} \rangle_{v_0}$};
      %\draw (-4.5,.1) node[below] (q0) {$\rho_1'$};
      \draw (1.3,.1) node[below] (q0) {$\rho_2'$};

      \draw (0,9) node[below] (q0) {$\alpha$};
      \draw (1,6) node[below] (q0) {$\beta$};
      \draw (-1.3,2) node[below] (q0) {$\gamma\rhobar$};
    \end{tikzpicture}
    \caption{Play $\rho$.}
    \label{fig:cycle}
  \end{minipage}
  \hfill
  \begin{minipage}[b]{0.48\linewidth}
    \centering
    \begin{tikzpicture}[yscale=.5,xscale=.5]
      \everymath{\scriptstyle}
      \draw (0,10) -- (-5,0);
      \draw (0,10) -- (5,0);

      \draw[fill=black] (0,10) circle (2pt);
      \draw[thick] (0,10) .. controls (-.5,8.5) .. (.5,7);

      \draw[fill=black] (-.2,8.1) circle (2pt);
      \draw[thick,dashed] (-.2,8.1) .. controls (-3,5) and (-1,2) .. (-4.5,0);
 
      \draw (-.5,1.5) node[below] (q0) {$\pi = \langle
        (\tau_i)_{\ipi} \rangle_{v_0}$};
      \draw (-4.5,.1) node[below] (q0) {$\pi_1'$};
      \draw (2.2,1.5) node[below] (q0) {$\pi_2'$};

      \draw (0,9) node[below] (q0) {$\alpha$};

\begin{scope}[xshift=1cm,yshift=3cm]
      \draw[white,fill=black!10!white] (-.5,4) -- (2,-1) -- (-3,-1);

      \draw[fill=black] (-.5,4) circle (2pt);
      \draw[thick] (-.5,4) .. controls (-1,3) and (-.5,2) .. (-1.5,0);
      \draw[thick,dotted] (-1.5,0) .. controls  (-1.7,-.75) ..  (-1.7,-1.5);

      \draw[fill=black] (-.825,2.3) circle (2pt);
      \draw[thick,dashed] (-.825,2.3) .. controls (1,.8)  .. (1.2,0);
      \draw[thick,dotted] (1.2,0) .. controls  (1.2,-.75) ..  (1.1,-1.5);
      \draw (-1.3,2) node[below] (q0) {$\gamma\rhobar$};
\end{scope}
    \end{tikzpicture}
    \caption{Play $\pi$ and possible deviations.}
    \label{fig:remove cycle}
  \end{minipage}
  \hfill\null
\end{figure}

  Assume that there exists a $\prec_j$-profitable deviation $\tau_j'$
  for player~$j$ w.r.t. $(\tau_i)_{\ipi}$. Let $\pi'$ be the outcome
  of the strategy profile $(\tau_j', \tau_{-j})$ from $v_0$, and
  $(y_1',\ldots,y_n')$ its cost profile.  Then we know that
  $(y_1,\ldots,y_n) \prec_j (y_1',\ldots,y_n')$. Two possible
  situations occur according to where player~$j$ deviates from
  $\pi$. We show that the first situation is impossible. In the second
  one, we construct a $\prec_j$-profitable deviation~$\sigma_j'$
  for player~$j$ w.r.t.~$(\sigma_i)_{\ipi}$, and then get a
  contradiction with $(\sigma_i)_{\ipi}$ being a secure equilibrium.
\begin{enumerate}[(1)]
\item[$(i)$] player~$j$ deviates from~$\pi$ strictly before depth
  $|\alpha|$ (see the play~$\pi'_1$ in Fig.~\ref{fig:remove cycle}).

  \smallskip
  Let us consider the prefix $h$ of $\pi'$ of length $|\alpha|$.  We
  first state that $h$ cannot visit $\F_j$ in a way that $\cost_j(h) <
  x_j$, because $h$ is consistent with $\sigma_{-j}$ (by definition of
  $(\tau_i)_{\ipi}$), and $(\sigma_i)_{\ipi}$ is a secure equilibrium.
  Therefore, $h$ is a $j$-\promising history
  w.r.t. $(\sigma_i)_{\ipi}$, as $\tau_j'$ is a $\prec_j$-profitable
  deviation w.r.t. $(\tau_i)_{\ipi}$.  By definition of
  $(\tau_i)_{\ipi}$, $\pi'$ is consistent with $\stropt_{-j}^v$ from
  $v=\last(h)$.  We consider the four possible cases of
  Lemma~\ref{lem:tec}:
  \begin{iteMize}{$\bullet$}
  \item $j \le k$. 
  
    \smallskip
    We have that $y_j = y'_j$.  The coalition $\Pi \setminus \{j\}$
    forces the play $\pi'$ to visit $\F_i$, for a certain $i > k$ (let
    us recall that $k < n$), before depth $|\alpha|+|V|$ as
    $\stropt_{-j}^v$ is memoryless. And so, $y_i' < |\alpha|+|V| \leq
    |\alpha| + |\gamma| \leq y_{k+1} \le y_i$ (as $|\alpha\beta\gamma|
    \le x_{k+1}$ and by Eq.~\eqref{eq:x_i>y_i}). This contradicts the
    fact that $(y_1,\ldots,y_n) \prec_j (y_1',\ldots,y_n')$.
    \smallskip
  \item $k < j \le l$.
  
    \smallskip
    The coalition $\Pi \setminus \{j\}$ prevents the play $\pi'$ from
    visiting $\F_j$, and so, $y_j'=+\infty$. As $y_j < +\infty$, it
    cannot be the case that $(y_1,\ldots,y_n) \prec_j
    (y_1',\ldots,y_n')$.
    \smallskip
  \item $l < j$ and $\cost(\rho_{\le |h|}) \preceq_j \cost(h)$.
  
    \smallskip
       The coalition $\Pi \setminus \{j\}$ forces the
    play $\pi'$ to visit $\F_i$, for a certain $i > k$, $i \neq j$,
    before depth $|\alpha|+|V|$, while avoiding the visit of $\F_j$
    (then, $y_j = y'_j=+\infty$).  As in the first case, this leads to
    a contradiction with the fact that $(y_1,\ldots,y_n) \prec_j
    (y_1',\ldots,y_n')$.
    \smallskip
   \item $l < j$ and $\cost(\rho_{\le |h|}) \not\preceq_j \cost(h)$.
      
       \smallskip
    Like in the second case, the coalition $\Pi \setminus \{j\}$
    prevents the play $\pi'$ from visiting $\F_j$, and so, $y_j =
    y'_j=+\infty$. Moreover, the hypothesis $\cost(\rho_{\le |h|})
    \not\preceq_j \cost(h)$ implies that $(y_1,\ldots,y_n) \prec_j
    (y_1',\ldots,y_n')$ cannot be true.
  \end{iteMize}

\smallskip

\item[$(ii)$] player~$j$ deviates from $\pi$ after depth $|\alpha|$
  ($\pi$ and $\pi'$ coincide at least on $\alpha$, see the
  play~$\pi'_2$ in Fig.~\ref{fig:remove cycle}).

  \smallskip
  We define for all histories $h \in H_j$:
  \[ \sigma_j'(h) := \left\{
  \begin{array}{ll}
  \sigma_j(h)
  & \mbox{ if $\alpha\beta \not\le h$,}\\
  \tau_j'(\alpha\delta) & \mbox{ if $h=\alpha\beta\delta$.}
\end{array}\right. \]
Let us set $\rho'=\langle \sigma_j',\sigma_{-j} \rangle_{v_0}$ of cost
profile $(x_1',\ldots,x_n')$. As player~$j$ deviates after $\alpha$
with the strategy $\tau_j'$, one can prove that
$$\pi'=\alpha\pibar' \quad \text{and} \quad \rho'=\alpha\beta\pibar'$$
by definition of $(\tau_i)_{\ipi}$ (see the play $\rho_2'$ in
Fig.~\ref{fig:cycle}). Since $\visit(\alpha)=\visit(\alpha\beta)$,
Equations~\eqref{eq:x_i=y_i fin},~\eqref{eq:x_i>y_i}
and~\eqref{eq:x_i=y_i} also stand by replacing $x_i$ with $x_i'$ and
$y_i$ with $y_i'$ (but the value of $l$ might be
different). Then $$(x_1,\ldots,x_n) \prec_j
(x_1',\ldots,x_n') \text{ iff } (y_1,\ldots,y_n) \prec_j
(y_1',\ldots,y_n'),$$ which proves that $\sigma_j'$ is a
$\prec_j$-profitable deviation for player~$j$
w.r.t. $(\sigma_i)_{\ipi}$, and this is a contradiction. 
\qed
\end{enumerate}
%%\end{proof}

\subsubsection*{\Goaldevopt secure equilibrium.}

\quad The next lemma uses the ideas developed in the proof of
Lemma~\ref{lem:remove a cycle} to show that any secure equilibrium can
be transformed into one that is \devopt. It is the last step before
proving Proposition~\ref{prop:goalopt devopt}, and finally Part $(ii)$
of Proposition~\ref{prop:se inf iff se fin}.

\begin{lem}\label{lem:es devopt}
  Let~$(\sigma_i)_{\ipi}$ be a secure equilibrium in a
  game~$(\mathcal{G},v_0)$, with outcome~$\rho$.
  Then there exists a \devopt secure equilibrium $(\tau_i)_{\ipi}$ in
  $(\mathcal{G},v_0)$ with outcome $\rho$.
\end{lem}

\proof
%%\begin{proof}
Let $\alpha$ be the prefix of $\rho$ of length $\max\{\cost_i(\rho)~|~
\cost_i(\rho) <+\infty\}$. It follows that $\visit(\rho) =
\visit(\alpha)$. Then we define the required strategy profile
$(\tau_i)_{\ipi}$ exactly like in the proof of Lemma~\ref{lem:remove a
  cycle}. We only remove the first line of the definition: $\tau_i(h)
= \sigma_i(\alpha\beta\delta)$ if $h=\alpha\delta$.  One can be
convinced that $(\tau_i)_{\ipi}$ and $(\sigma_i)_{\ipi}$ have the same
outcome $\rho$. We prove in the exact same way that $(\tau_i)_{\ipi}$
is a secure equilibrium in $(\mathcal{G},v_0)$ (here, $k=l$).

Let us now show that $(\tau_i)_{\ipi}$ is \devopt\ thanks to
Lemma~\ref{lem:devopt}.  Let $\tau_j'$ be a strategy of some
player~$j$ such that the play $\rho'=\langle \tau_j', \tau_{-j}
\rangle_{v_0}$ verifies
  \begin{enumerate}[(i)]
    \item $\cost_j(\rho) = \cost_j(\rho')$,
    \item $\forall \, i \in \Pi$ such that $\cost_i(\rho) < +
      \infty$, we have that $\cost_i(\rho) \le \cost_i(\rho')$,
    \item $\exists\, i \in \Pi \ \cost_i(\rho) < \cost_i(\rho')$.
  \end{enumerate}
  We must prove that there exists $l$ such that
  $\cost_l(\rho)=+\infty$ and $\cost_l(\rho') \leq \depthdev =
  \max\{\cost_i(\rho)~|~ \cost_i(\rho) <+\infty\} + |V|$.  Notice that
  $\cost(\rho)=\cost(\alpha)$.
  
  On one hand, suppose that $\cost(\alpha) \not\prec_j
  \cost(\rho'_{\le |\alpha|})$.  By ($i$), ($ii$) and ($iii$), the
  only possibility is to have some $l$ such that
  $\cost_l(\alpha)=+\infty$ and $\cost_l(\rho'_{\le
    |\alpha|})<+\infty$, that is, $\cost_l(\rho)=+\infty$ and
  $\cost_l(\rho') \le |\alpha| < \depthdev$.
  
  On the other hand, if $\cost(\alpha) \prec_j \cost(\rho'_{\le
    |\alpha|})$, then according to the last case of
  Definition~\ref{def:promising}, $\rho'_{\le |\alpha|}$ is
  $j$-\promising w.r.t. $(\sigma_i)_{\ipi}$. Indeed, $\rho'_{\le
    |\alpha|}$ is consistent with $\sigma_{-j}$, and there
  exists $\ipi$ such that $\cost_i(\rho)=+\infty$ (otherwise it would
  contradict the fact that $(\sigma_i)_{\ipi}$ is a secure
  equilibrium). By definition of $(\tau_i)_{\ipi}$, $\rho'$ is thus
  consistent with $\stropt_{-j}^v$ from vertex $v =
  \rho'_{|\alpha|}$. Thus, by Lemma~\ref{lem:tec} (first case or third
  case), there exists $l$ such that $\cost_l(\rho)=+\infty$ and
  $\cost_l(\rho') < |\alpha| + |V| = \depthdev$ (as $\stropt_{-j}^v$
  is memoryless).

  In both cases, by Lemma~\ref{lem:devopt}, we proved that
  $(\tau_i)_{\ipi}$ is \devopt.  
  \qed
%%\end{proof}

\noindent We are now able to prove Proposition~\ref{prop:goalopt devopt}, which
states that if there exists a secure equilibrium in a game
$(\mathcal{G},v_0)$, then there exists one which is \goalopt and
\devopt.

\proof[Proof of Proposition~\ref{prop:goalopt devopt}]
%%\begin{proof}[of Proposition~\ref{prop:goalopt devopt}]
  Let $(\sigma_i)_{\ipi}$ be a secure equilibrium in $(\mathcal{G},v_0)$ with
  outcome $\rho=\langle (\sigma_i)_{\ipi} \rangle_{v_0}$ and cost profile
  $(x_i)_{\ipi}$. Let us assume w.l.o.g. that $\Pi=\{1,\ldots,n\}$ and
  $$x_1 \leq \ldots \leq x_{l} < x_{l+1}=\ldots=x_n=+\infty$$ where $0
  \leq l \leq n$. Let us set $x_0=0$. For all $k \in
  \{0,1,\ldots,l-1\}$ such that $(x_{k+1}-x_k) \geq 2\cdot|V|$ and
  while it is still the case, we apply the following procedure to get
  a \goalopt secure equilibrium.

  Consider such a $k \in \{0,1,\ldots,l-1\}$. 
  Then, we can write $\rho=\alpha\beta\gamma\rhobar$, with
  $\beta$ non-empty, $|\gamma|
  \ge |V|$, and such that
  \begin{align*}
    & x_k \leq |\alpha\beta\gamma| \leq x_{k+1}\\
    & \visit(\alpha)=\visit(\alpha\beta\gamma)=\{1,\ldots,k\}\\ 
    & \last(\alpha)=\last(\alpha\beta).
  \end{align*}
  Let us remark that $\visit(\rho) \not= \visit(\alpha)$ as $k < l$.
  By Lemma~\ref{lem:remove a cycle} there exists a secure equilibrium
  in $(\mathcal{G},v_0)$ with outcome $\alpha\gamma\rhobar$. Its cost
  profile $(y_i)_{\ipi}$ is such that
  \begin{align*}
    & y_1=x_1,\ldots,y_{k}=x_{k};\\
    & y_{k+1}<x_{k+1},\ldots,y_{l}<x_{l};\\
    & y_{l+1}=x_{l+1}=+\infty,\ldots,y_{n}=x_{n}=+\infty.
    \end{align*}
  By applying finitely many times this procedure, we can assume
  w.l.o.g. that $(\sigma_i)_{\ipi}$ is a secure equilibrium with a cost
  profile $(x_1,\ldots,x_n)$ such that
  \[\begin{array}{ll}
  x_i < i\cdot 2 \cdot |V| & \text{\ \ for } i \leq l\\
  x_i = +\infty & \text{\ \ for } i > l,
  \end{array}\]
  meaning that $(\sigma_i)_{\ipi}$ is a \goalopt secure equilibrium.
  
  Moreover, by Lemma~\ref{lem:es devopt}, there exists a
  \devopt secure equilibrium with the same outcome, i.e. a
  \goalopt and \devopt secure equilibrium. And this concludes the
  proof.  
  \qed
%%\end{proof}

\begin{rem}\label{rem:goaloptdevopt costs}
  Regarding the costs, this proof shows that if there exists a secure
  equilibrium with cost profile $(a_i)_{i \in \Pi}$ in a game
  $(\mathcal{G},v_0)$, then there exists a \goalopt and \devopt secure
  equilibrium with cost profile $(b_i)_{i \in \Pi}$ in $(\mathcal{G},v_0)$,
  such that for all $i \in \Pi$, $b_i \le a_i$. In particular, the
  cost profile is usually not preserved.
\end{rem}

Finally, on the basis of Proposition~\ref{prop:goalopt devopt}, we are
able to prove Part ($ii$) of Proposition~\ref{prop:se inf
  iff se fin}: given a game $(\mathcal{G},v_0)$, if there exists a secure
equilibrium in~$(\mathcal{G},v_0)$, then there exists a \goalopt and
\devopt secure equilibrium in $\ttree{\depth}$, for $\depth =
\depthgoal + 3\cdot |V|$.

\proof[Proof of Proposition~\ref{prop:se inf iff se fin}, Part ($ii$)]
%%\begin{proof}[of Proposition~\ref{prop:se inf iff se fin}, Part ($ii$)]
  Let $(\sigma_i)_{\ipi}$ be a secure equilibrium in $(\mathcal{G},v_0)$ with
  outcome $\rho$.  By Proposition~\ref{prop:goalopt devopt}, we can
  suppose w.l.o.g. that $(\sigma_i)_{\ipi}$ is \goalopt and \devopt.
  Let us define the strategy profile $(\tau_i)_{\ipi}$ in
  $\ttree{\depth}$ as the strategy profile $(\sigma_i)_{\ipi}$
  restricted to the finite tree $\ttree{\depth}$. We prove that
  $(\tau_i)_{\ipi}$ is a secure equilibrium in $\ttree{\depth}$, which
  is clearly \goalopt ($d > \depthgoal$).
  
  For a contradiction, assume that player~$j$ has a
  $\prec_j$-profitable deviation $\tau'_j$
  w.r.t. $(\tau_i)_{\ipi}$. Let us denote $\pi = \langle
  (\tau_i)_{\ipi} \rangle_{v_0}$ and $\pi' = \langle \tau_j',\tau_{-j}
  \rangle_{v_0}$ in $\ttree{\depth}$.  We extend arbitrarily $\tau'_j$ in
  $\ginf$, into a strategy denoted $\sigma'_j$, and let $\rho' =
  \langle \sigma_j',\sigma_{-j} \rangle_{v_0}$. Let us remark that $\pi$
  (resp. $\pi'$) is a prefix of $\rho$ (resp. $\rho'$) of length
  $\depth > \depthgoal$, and thus, in particular $\cost(\rho)=
  \cost(\pi)$. Moreover, it is impossible that $\cost_j(\pi) >
  \cost_j(\pi')$, otherwise we would have $\cost_j(\rho) >
  \cost_j(\rho')$ and so, get a contradiction with the fact that
  $(\sigma_i)_{\ipi}$ is a secure equilibrium in $\ginf$.  Then,
  player~$j$ gets the same cost $\cost_j(\pi) = \cost_j(\pi')$ and
  $$
  \forall \ipi \ \cost_i(\pi) \le \cost_i(\pi') 
 \quad  \wedge \quad \exists \ipi \ \cost_i(\pi) < \cost_i(\pi').$$
 %
  % Let $h$ be the prefix of $\pi'$ of length $\cost_k(\pi)$.
  % As $\tau'_j$ is $\prec_j$-profitable, we have
  % $$\cost(\pi) \prec_j \cost(h).$$ 

 \noindent We now show that $\cost_j(\rho)=\cost_j(\rho')$. In the case where
 $\cost_j(\pi) = \cost_j(\pi') =+\infty$ ($=\cost_j(\rho)$), we must
 have $\cost_j(\rho') = +\infty$. Otherwise, it would contradict the
 fact that $(\sigma_i)_{\ipi}$ is a secure equilibrium in $\ginf$.  In
 the case where $\cost_j(\pi) = \cost_j(\pi') <+\infty$, then
 $\cost_j(\rho)=\cost_j(\rho')$ (as $\pi$ and $\pi'$ are prefixes of
 $\rho$ and $\rho'$ respectively). Moreover, since $\tau'_j$ is a
 $\prec_j$-profitable deviation w.r.t. $(\tau_i)_{\ipi}$, it follows
 that for all $i \in \Pi$ such that $\cost_i(\rho) < + \infty$, we
 have that $\cost_i(\rho) \le \cost_i(\rho')$, and there exists $i \in
 \Pi$ such that $\cost_i(\rho) < \cost_i(\rho')$.  As
 $(\sigma_i)_{\ipi}$ is \devopt, Lemma~\ref{lem:devopt} implies that
 there exists some $l \in \Pi$ such that $\cost_l(\rho)=+\infty$ and
 $\cost_l(\rho') < \depthdev=\max\{\cost_i(\rho)~|~ \cost_i(\rho)
 <+\infty\} + |V|$.  As $\depthdev \leq \depthgoal + |V| < \depth$, we
 have that $\cost_l(\pi)=\cost_l(\rho)=+\infty$ and $\cost_l(\pi') =
 \cost_l(\rho') < \depthdev$.  This gives a contradiction with the
 fact that $\tau'_j$ is a $\prec_j$-profitable deviation
 w.r.t. $(\tau_i)_{\ipi}$ in $\ttree{\depth}$. Therefore,
 $(\tau_i)_{\ipi}$ is a secure equilibrium in this game. On the other
 hand, the previous argument also shows that $(\tau_i)_{\ipi}$ is
 \devopt. 
 \qed
%%\end{proof}

\begin{rem}\label{rem:(ii) costs}
  This proof shows in particular that if there exists a \goalopt and
  \devopt secure equilibrium in~$(\mathcal{G},v_0)$, then there exists a
  \goalopt and \devopt secure equilibrium in $\ttree{\depth}$
  \emph{with the same cost profile}. Together with
  Remark~\ref{rem:goaloptdevopt costs}, we then proved the following
  result: if there exists a secure equilibrium with cost profile
  $(a_i)_{i\in\Pi}$ in $(\mathcal{G},v_0)$, then there exists a \goalopt and
  \devopt secure equilibrium with cost profile $(b_i)_{i\in\Pi}$ in
  $\ttree{\depth}$, such that for all $i \in \Pi$, $b_i \le a_i$.
\end{rem}

Remarks~\ref{rem:(i) costs} and~\ref{rem:(ii) costs} imply the
proposition below.

\begin{prop}\label{prop:dec pbm}
  Given a multiplayer quantitative reachability game and a tuple of
  thresholds $(t_i)_{i\in\Pi} \in ({\mathbb R} \cup
  \{+\infty\})^{\Pi}$, one can decide in {\sf ExpSpace} whether there
  exists a secure equilibrium with cost profile $(c_i)_{i\in\Pi}$ such
  that for all $i \in \Pi$, $c_i \le t_i$.
\end{prop}
The decision problem related to Proposition~\ref{prop:dec pbm} is
equivalent to decide whether there exists a \goalopt and \devopt
secure equilibrium with cost profile $(a_i)_{i\in\Pi}$ in
$\ttree{\depth}$ where $\depth=\depthgoal + 3\cdot |V|$, such that for
all $i \in \Pi$, $a_i \le t_i$.  Notice that $\depth$ does not depend
on $(t_i)_{i\in\Pi}$.

\section{Conclusion and Perspectives}

In this paper, we study the concept of subgame perfect equilibrium, a
refinement of Nash equilibrium well-suited to the framework of games
played on graphs. We also introduce the new concept of subgame
perfect secure equilibrium. We prove the existence of subgame perfect
equilibria in multiplayer quantitative reachability games. We also
prove the existence of subgame perfect secure equilibria, but only in
the two-player framework. Finally, we provide an algorithm deciding
in {\sf ExpSpace} the existence of secure equilibria in the
multiplayer case. On the one hand, the first two results have been
obtained by topological techniques, that are completely different from
the techniques used in~\cite{BBD10,BBDlong}.  On the other hand,
proofs of the last result are strongly inspired by proofs developed in
these references, but have required new ideas about the coalition
strategies. 

%however with new ideas about the coalition strategies.
  
There are several interesting directions for future research. We are
currently working on the model of quantitative game, enriched by
allowing $n$-tuples of positive weights on edges (see
Theorem~\ref{theo:ex spe tuple of costs}). We do believe that our
results remain true in this context. The case of Nash equilibria is
already treated in~\cite{BBDlong}.  Notice that our results trivially
generalize to the particular case where the weights of the edges are
of the form $(c, \ldots, c)$ with $c \in \IN_0$. Indeed it is enough
to replace each such edge by a path of length $c$ composed of $c$ new
edges (of cost 1).

To the best of our knowledge, the existence of secure equilibria in
the multi-player framework is still an open problem. We prove that the
existence of a secure equilibrium in an infinite game is equivalent to
the existence of a \goalopt and \devopt secure equilibrium in a
finite game. This open problem could be positively solved if
Corollary~\ref{coro:kuhn secure} could be adapted in a way to get a
\goalopt and \devopt secure equilibrium in the finite game, and then
by applying Proposition~\ref{prop:se inf iff se fin}. A deeper
understanding of equilibria with unnecessary cycles could also be
helpful. For the moment, we are not able to solve this problem with
more than two players. The same kind of question is also open for
subgame perfect secure equilibria.

Another research direction concerns a deeper study of the memory
needed in the different kinds of equilibria. In the case of subgame
perfect equilibria and subgame perfect secure equilibria, the
topological techniques give no results on the memory needed. However,
in the case of secure equilibria, we prove that we can limit to
finite-memory equilibria.

\section*{Acknowledgements}
  This work has been partly supported by a grant from the National
  Bank of Belgium, the ARC project (number AUWB-2010-10/15-UMONS-3),
  and the ESF project GASICS. The third author is supported by a grant
  from F.R.S.-FNRS. The authors wish to thank the anonymous reviewers
  for their useful comments and pointing out \cite{har85,FL83}.

\bibliographystyle{alpha}
\bibliography{main}

\end{document}